\documentclass[aps,pra,reprint,superscriptaddress,nofootinbib]{revtex4-2}
\usepackage[utf8]{inputenc}
\usepackage{amsmath}
\usepackage{amssymb}
\usepackage{physics}
\usepackage{bbold}
\usepackage{graphicx}
\usepackage{multirow}
\usepackage{hyperref}
\hypersetup{colorlinks=true, citecolor=blue}
\usepackage{cleveref}
\usepackage{xcolor}
\usepackage{bbold} 
\usepackage{tabularx} 
\usepackage{soul}  

\usepackage[caption = false]{subfig}

\definecolor{pink}{HTML}{F282B4}

\newcommand{\KL}{D_{\text{KL}}}
\newcommand{\nbins}{n_{\text{bins}}}
\newcommand{\rhot}{\rho_{\mathbf{\theta}}}
\newcommand{\Ht}{H_{\mathbf{\theta}}}
\newcommand{\qt}{q_{\mathbf{\theta}}}
\newcommand{\pt}{|p_\text{T}^{\text{rel}}|}

\usepackage{makecell} 
\usepackage{mathtools}
\usepackage{tabularx} 
\usepackage{amsmath}
\usepackage{amssymb}

\begin{document}

\title{Learning to generate high-dimensional distributions\\ with low-dimensional quantum Boltzmann machines}

\author{Cenk T\"uys\"uz}
\email{cenk.tueysuez@desy.de}
\affiliation{CQTA, Deutsches Elektronen-Synchrotron DESY, Platanenallee 6, 15738
Zeuthen, Germany}
\affiliation{Institut für Physik, Humboldt-Universit\"at zu Berlin,
Newtonstr. 15, 12489 Berlin, Germany}
\author{Maria Demidik}
\affiliation{CQTA, Deutsches Elektronen-Synchrotron DESY, Platanenallee 6, 15738
Zeuthen, Germany}
\affiliation{Computation-Based Science and Technology Research Center, The Cyprus Institute, 20 Kavafi Street, 2121 Nicosia, Cyprus}
\author{Luuk Coopmans}
\affiliation{Quantinuum, Partnership House, Carlisle Place, London SW1P 1BX, United Kingdom}
\author{Enrico Rinaldi}
\affiliation{Quantinuum K.K., Otemachi Financial City Grand Cube 3F, 1-9-2 Otemachi, Chiyoda-ku, Tokyo, Japan}
\affiliation{Interdisciplinary Theoretical and Mathematical Sciences (iTHEMS) Program, RIKEN, Wako, Saitama 351-0198, Japan}
\author{Vincent Croft}
\affiliation{Applied Quantum Algorithms Leiden $\left<aQa^L\right>$ and Leiden Institute of Advanced Computer Science (LIACS), Leiden University, Niels Bohrweg 2, 2333 CA Leiden, Netherlands
}
\author{Yacine Haddad}
\affiliation{Department of Physics, Northeastern University, Boston, MA 02115 USA}
\author{Matthias Rosenkranz}
\affiliation{Quantinuum, Partnership House, Carlisle Place, London SW1P 1BX, United Kingdom}
\author{Karl Jansen}
\affiliation{Computation-Based Science and Technology Research Center, The Cyprus Institute, 20 Kavafi Street, 2121 Nicosia, Cyprus}
\affiliation{CQTA, Deutsches Elektronen-Synchrotron DESY, Platanenallee 6, 15738
Zeuthen, Germany}

\begin{abstract}
In recent years, researchers have been exploring ways to generalize Boltzmann machines (BMs) to quantum systems, leading to the development of variations such as fully-visible and restricted quantum Boltzmann machines (QBMs). Due to the non-commuting nature of their Hamiltonians, restricted QBMs face trainability issues, whereas fully-visible QBMs have emerged as a more tractable option, as recent results demonstrate their sample-efficient trainability. These results position fully-visible QBMs as a favorable choice, offering potential improvements over fully-visible BMs without suffering from the trainability issues associated with restricted QBMs. In this work, we show that low-dimensional, fully-visible QBMs can learn to generate distributions typically associated with higher-dimensional systems. We validate our findings through numerical experiments on both artificial datasets and real-world examples from the high energy physics problem of jet event generation. We find that non-commuting terms and Hamiltonian connectivity improve the learning capabilities of QBMs, providing flexible resources suitable for various hardware architectures. Furthermore, we provide strategies and future directions to maximize the learning capacity of fully-visible QBMs.\looseness=-1
\end{abstract}

\maketitle

\section{Introduction}

Generative learning has garnered widespread attention in classical machine learning in recent years~\cite{goodfellow_generative_2014, lampinen_bayesian_2001, kingma_auto-encoding_2022, diff_models}. Generative models have found applications across a range of fields, from drug discovery~\cite{korshunova_generative_2022} to forecasting the dynamics of high-dimensional complex systems~\cite{gao_generative_2024}. The core idea behind generative models is to learn the joint probability distribution of a data set. Once trained, these models can generate new data samples drawn from the learned distribution.

With the emergence of quantum machine learning (QML), many classical generative models have been adapted to the quantum framework to leverage the potential advantages of quantum computing. QML models hold the promise of leveraging greater expressive power and better generalization as expressed in Refs.~\cite{perdomo-ortizOpportunitiesChallengesQuantumassisted2018,benedettiParameterizedQuantumCircuits2019}. Some of the popular QML models for generative learning are quantum generative adversarial networks~\cite{quantum_GAN_2018}, quantum variational autoencoders~\cite{khoshaman_quantum_VAE_2018} and quantum circuit Born machines~\cite{PhysRevA.98.062324, coyle_born_2020, benedettiParameterizedQuantumCircuits2019}. These quantum models face significant challenges related to scalability and trainability~\cite{mccleanBarrenPlateausQuantum2018,ortizmarreroEntanglementInducedBarrenPlateaus2021}, particularly when deployed on current quantum hardware~\cite{preskill_nisq_quantum_2018, benedettiGenerativeModelingApproach2019, wang_noise-induced_2021, you_exponentially_2021, tuysuz_symmetry_2024, ragone_lie_2024, rudolph_trainability_2023, crognaletti2024estimates}.

An alternative QML model is the quantum Boltzmann machine (QBM)~\cite{amin_quantum_2018,benedettiQuantumAssistedLearningHardwareEmbedded2017,kieferova_tomography_2017,denilImplementationQuantumRBM2011}, which is a generalization of the classical Boltzmann machine (BM)~\cite{ackleyLearningAlgorithmBoltzmann1985,hinton_practical_2012}. BM is a probabilistic network of binary units with an associated energy function described by a classical spin Hamiltonian~\cite{glauberTimeDependentStatisticsIsing1963,sherringtonSolvableModelSpinGlass1975}. In general, there are two types of units: visible and hidden. Visible units correspond to observed variables given by input and/or output, while hidden units are not observed. A BM is called a fully-visible model if it only consists of visible units. Fully-visible models are known to have poor expressivity~\cite{montufar_expressive_2014} and therefore are not widely used. Instead, restricted Boltzmann machines (RBMs) are frequently used in the literature due to being universal approximators~\cite{le_roux_representational_2008}. The expressivity of RBMs increases with the number of hidden units~\cite{le_roux_representational_2008, montufar_expressive_2014}. However, exactly evaluating the partition function of RBMs is computationally intractable as the system size increases. For this reason, there have been extensive efforts to use approximate methods to implement them~\cite{pmlr-vR5-carreira-perpinan05a}. Nevertheless, RBMs are still considered difficult to scale and practical implementations do not exceed hundreds of units~\cite{long_restricted_2010}.

\begin{figure*}[!t]
    \centering
    \includegraphics[width=\linewidth]{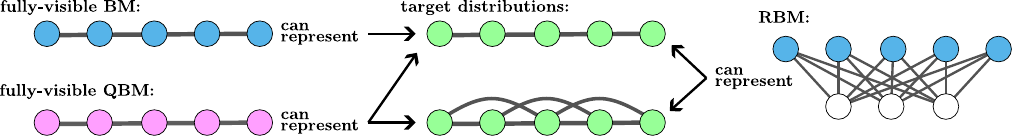}
    \caption{\textbf{Summary of the main result.} Fully-visible BMs can generate distributions that match their connectivity. A nearest-neighbor connected BM can represent a distribution generated by nearest-neighbor statistics. We show that a fully-visible QBM, on the other hand, can represent distributions generated by higher dimensional models. Alternatively, a classical RBM, consisting of visible (blue) and hidden (hollow, white) units, can also learn both of these distributions presented; however, it requires additional hidden units for this purpose.}
    \label{fig:qbm-result}
\end{figure*}

Quantum computing holds the promise of efficiently sampling from the Gibbs distribution corresponding to BMs and RBMs, which can potentially allow scaling of BM models~\cite{PhysRevLett.103.220502, temme_quantum_2011}. Preparing the Gibbs state on a quantum computer also allows generalizations to non-commuting and off-diagonal Hamiltonians~\cite{chenQuantumThermalState2023,chenEfficientExactNoncommutative2023}. While recent results suggest the classical simulability of high temperature Gibbs states~\cite{bakshi2024hightemperaturegibbsstatesunentangled}, preparing low temperature Gibbs states still requires quantum computers~\cite{rouzé2024efficientthermalizationuniversalquantum}. These recent developments open the possibility to generalize BMs to QBMs.

Similar to their classical counterparts, QBMs can also be implemented as fully-visible and restricted models. Restricted QBMs, being the more studied of the two types, have been shown to suffer from trainability issues and inefficient gradient evaluation~\cite{amin_quantum_2018, wiebe_generative_2019, kieferova_tomography_2017}. Several methods have been proposed in the QML literature to address the challenges associated with training QBMs. Some approaches impose restrictions on the Hamiltonian terms~\cite{amin_quantum_2018, wiebe_generative_2019}, while other approaches are not scalable~\cite{zoufal_variational_2021, huijgen_training_2024}. In contrast, recent results suggest that fully-visible QBMs can be trained using a polynomial number of Gibbs states~\cite{coopmans_sample_2024, patel2024quantum}. This leads to the natural question of whether fully-visible QBMs can be expressive models in comparison to inexpressive fully-visible BMs in practice. 

In this work, we extend the existing literature on quantum generative learning and highlight the advantages of fully-visible QBMs. First, we demonstrate that fully-visible QBMs are capable of learning higher-dimensional probability distributions even with limited connectivity. Here, we define the dimension of a probability distribution as the dimension of the lattice that describes the interaction of each binary variable. We illustrate this result in Fig.~\ref{fig:qbm-result}. Second, we address the role of the Hamiltonian in enhancing the model's expressivity. While previous studies considered the transversal field Ising Hamiltonian and spin-glass Hamiltonian~\cite{amin_quantum_2018, kappen_learning_2020}, our experiments highlight that more general Hamiltonians can boost QBM learning capacity with negligible computational overhead~\cite{coopmans_sample_2024}. Finally, we showcase the applicability of our findings to real-world examples in learning to generate reduced-size particle jet events. 

Particle jets are clusters of particles that are observed at particle collider experiments such as the Large Hadron Collider~\cite{butter_machine_2023}. Jets can originate from quarks or gluons, and studying them provides a fundamental understanding of the Standard Model of particle physics and beyond~\cite{cms_collaboration_search_2020}. Simulating jet events is essential to testing existing and new theories, but the computational cost of this task is a limiting factor~\cite{Jordan:2011ci, ai-hep}. These simulations are often performed using computationally demanding Monte Carlo methods and require simulating billions of events that model the interaction of particles with the detector material~\cite{10.21468/SciPostPhysCodeb.8, Bellm:2015jjp}. In recent years, researchers have employed classical generative deep learning techniques such as graph neural networks (GNNs)~\cite{butter_machine_2023} to overcome this problem. Despite their early success, these methods are not able to learn the correlations between particles accurately~\cite{kansal_particle_2021}.

Moreover, although detector measurements at collider experiments are considered classical data, recent experiments suggest that quantum entanglement can be detected through these measurements~\cite{aad_observation_2024, cms2024-1, cms2024-2}. These results motivate the growing interest to approach this generative learning problem using QML methods~\cite{delgado_quantum_2022, delgado_quantum_2022b, Borras_2023, di_meglio_quantum_2023}. However, most prior work either uses methods such as feature reduction techniques that do not capture the high-order correlations or variational methods that were shown to suffer from issues such as barren plateaus~\cite{delgado_unsupervised_2022, rousselot_generative_2023, rudolph_trainability_2023}. In this work, we apply QBMs to the particle jet event generation problem for the first time.

The paper is organized as follows. Section~\ref{sec:framework} consists of model definitions, algorithms, numerical methods and mathematical tools used to obtain the results. In Sec.~\ref{sec:qbm} we give the necessary definitions to construct BM and QBM models. Following that, we describe how to train these models. In Section~\ref{sec:tpq}, we describe the numerical tools we use and in Section~\ref{sec:mut-info}, we define concepts from information theory. We present our numerical results in Section~\ref{sec:results}. In Section~\ref{sec:exp-boltzmann}, we show results for target distributions that are artificially constructed, while in Section~\ref{sec:jet}, we provide results for the particle jet event generation problem using BMs and QBMs up to 16 qubits. In Section~\ref{sec:conclusion}, we conclude with a brief discussion of the limitations of this study and provide future directions. 
 
\section{\label{sec:framework} Framework}

\subsection{\label{sec:qbm} Fully-visible (quantum) Boltzmann machines}

\subsubsection{\label{sec:model} Model description}

A Boltzmann machine is described by the Gibbs state $\rhot$ of a classical or quantum Hamiltonian $\Ht$, parametrized by a set of parameters $\mathbf{\theta}$, at a finite inverse temperature $\beta=1/kT$,
\begin{equation}
\rhot = \frac{e^{-\beta \Ht}}{\mbox{Tr}(e^{-\beta \Ht})},
\label{eq:gibbs}
\end{equation}
and,
\begin{equation}
\Ht = \sum_{i} \theta_i H_i,
\label{eq:hamiltonian_definition}
\end{equation}
where $\forall\, i, \theta_i \in \mathbb{R}$ and $\mathbf{\theta}$ is the parameter vector that parametrizes each $H_i$, which are bounded Hermitian operators. In this work, we consider each $H_i$ to be a Pauli string with length $n$, excluding the identity string ($I^{\otimes n}$). A length $n$ Pauli string is a tensor product of $n$ Pauli matrices, e.g. $\sigma_0^X \sigma_1^I \sigma_2^Y$ is a length-$3$ Pauli string acting on qubits $0, 1, 2$. We use the shorthand notation $\sigma_0^X \sigma_0^I \sigma_2^Y \coloneqq \sigma^X \otimes \sigma^I \otimes \sigma^Y$. For simplicity, we often omit tensor product factors involving the identity.

The goal of the algorithm is to approximate the target state $\eta$ which embeds a classical probability distribution $p(s)$ with the output probability distribution $q_\theta(s)$ of $\rhot$. Here $s$ is a bit string of length $n$. If the Hamiltonian is diagonal in the computational basis, it is denoted as a classical Hamiltonian. In this case, the density matrix $\rhot$ is diagonal in the computational basis and it is a mixed state ($\Tr(\rhot)=1, \Tr(\rhot^2) \leq 1$). We denote such a model as ``BM''. If the Hamiltonian contains off-diagonal terms in the computational basis, we denote it as a quantum Hamiltonian. The density matrix of the quantum Hamiltonian contains non-zero off-diagonal entries in the computational basis. We denote such a model as ``QBM''. 

Typically, BMs are constructed with up to two body interactions, such that the corresponding Hamiltonian is a two-local classical Hamiltonian. This formulation has been widely adopted in the classical machine learning community~\cite{hinton_practical_2012}. Let us define the Hamiltonians that describe BMs and QBMs. Consider a lattice with connectivity defined by an undirected graph $\mathcal{G} (\mathcal{V}, \mathcal{E})$, where $\mathcal{V}$ represents the lattice sites and $\mathcal{E}$ denotes their connectivity. We define the system size $n$ as the number of lattice sites. Then, any two-local Hamiltonian can be expressed in the Pauli basis as follows:
\begin{equation}
    H_\theta = \sum_{k \in \mathcal{P}_1} \sum_{i \in \mathcal{V}} \theta_i^k \sigma_i^k + \sum_{(k,l) \in \mathcal{P}_2} \sum_{(i,j) \in \mathcal{E}} \theta_{i,j}^{k,l} \sigma_i^k \sigma_j^l,
\label{eq:two-local-hamiltonian}
\end{equation}
where $\sigma_i^k$ denotes the Pauli matrix applied on the $i$-th qubit with $k \in \mathcal{W}$ determining its type ($\mathcal{W}=\{ X, Y, Z \}$) and $\mathcal{P}_1 \subseteq \mathcal{W}$, $\mathcal{P}_2 \subseteq \mathcal{W} \otimes \mathcal{W}$ are sets of one- and two-local Pauli matrix types. Note that this formulation can be extended to high-weight Pauli strings, but we restrict the models up to two-body interactions throughout this work for practical reasons. 

A BM can be described with $\mathcal{P}_1 = \{ Z \}, \mathcal{P}_2 = \{ ZZ \}$. Common choices for QBM Hamiltonians contain physics inspired sets of operators, \textit{e.g.}, the transversal field Ising model with $\mathcal{P}_1 = \{ X, Z \}$ and $\mathcal{P}_2 = \{ ZZ \}$. The authors of Ref.~\cite{kappen_learning_2020} propose using $\mathcal{P}_1 = \{ X, Y, Z \}, \mathcal{P}_2 = \{ XX, YY, ZZ \}$ and report results, outperforming classical BMs. We introduce additional Hamiltonian definitions as presented in Table~\ref{tab:hamiltonian-defs}.

 \begin{table}[!h]
    \centering
    \caption{Two-local Hamiltonian definitions.}
    \label{tab:hamiltonian-defs}
    \begin{tabularx}{\linewidth}{X|l|X}
         Label & $\mathcal{P}_1$ & $\mathcal{P}_2$\\
         \hline
         \hline
         ising & $Z$ & $ZZ$\\
         \hline
         tfim & $X, Z$ & $ZZ$\\
         \hline
         spin-glass & $X, Y, Z$ & $XX, YY, ZZ$\\
         \hline
         spin-glass-real & $X, Z$ & $XX, YY, ZZ$\\
         \hline
         generic & $X, Y, Z$ & $XX, XY, XZ, YX$ \newline $YY, YZ, ZX, ZY, ZZ$ \\
         \hline
         generic-real & $X, Z$ & $XX, XZ, YY, ZX, ZZ$\\
         \hline
    \end{tabularx}
\end{table}

In this work, we consider only the fully-visible case, where all lattice sites are visible units. It is common for fully-visible models to have a connectivity graph that is a complete graph. 

Consider the target probability distribution $p(s)$ such that $\sum_s p(s) = 1$, where we sum over all possible bit strings of length $n$. The fully-visible BM encodes the target distribution into a density matrix such that
\begin{equation}
    \eta = \mbox{diag}(p(s)). 
\label{eq:diag_encoding}
\end{equation}

Leveraging the encoding in Eq.~\eqref{eq:diag_encoding} for QBMs, results in a mismatch between the model $\rhot$ and the target $\eta$. This occurs due to the non-zero off-diagonal entries present in the Gibbs state of a quantum Hamiltonian. To overcome this, Kappen~\cite{kappen_learning_2020} proposed the following encoding in the computational basis: 
\begin{equation}
\eta = \ket{\psi}\bra{\psi}, \quad  \ket{\psi} = \sqrt{p(s)}e^{i\alpha(s)}\ket{s},  
\label{eq:qbm-target}
\end{equation}
where $\alpha(s)$ is an arbitrary phase that can be chosen freely. We choose $\forall s, \alpha(s)=0$ for simplicity. Notice that this embedding results in a pure target state ($\Tr(\rho)=1, \Tr(\rho^2) = 1$), in contrast to the mixed target state encoding of the diagonal BM model.

\subsubsection{\label{sec:trainining} Training (quantum) Boltzmann machines}

We train BMs and QBMs using the quantum relative entropy as the loss function, which is defined as
\begin{equation}
    S(\eta \, || \, \rhot) = \mbox{Tr}(\eta \log \eta) - \mbox{Tr}(\eta \log \rhot),
\label{eq:quantum-rel-entropy}
\end{equation}
where $\eta$ is the target density matrix and $\rhot$ is the density matrix of the model. The first term on the right-hand side of Eq.~\eqref{eq:quantum-rel-entropy} corresponds to the negative von~Neumann entropy of $\eta$ ($S(\eta) = - \mbox{Tr}(\eta \log \eta)$) and the second term corresponds to the negative quantum log-likelihood between target and model density matrices. Plugging the target and model density matrices for BM into Eq.~\eqref{eq:quantum-rel-entropy} yields the Kullback-Leibler divergence ($\KL$),
\begin{align}
    \KL &(p(s) \, || \, \qt (s)) = -\sum_s p(s) \log \left( \frac{\qt(s)}{p(s)} \right) \nonumber \\
    &= \underbrace{-\sum_s p(s) \log \qt(s)}_{\text{negative~log-likelihood}} + \underbrace{\sum_s p(s) \log p(s)}_{\text{negative Shannon~entropy}},
\end{align}
where $p(s)$ denotes the target and $\qt(s) = \text{diag}(\rhot)$ denotes the model probability density for given bit string $s$. Therefore, one can minimize the quantum relative entropy in order to minimize $\KL$ between the target and model for both BM and QBM.

Quantum relative entropy can be minimized via gradient descent. This requires computing the gradients of all parameters of the Hamiltonian with respect to quantum relative entropy. Then, the gradients take the form,
\begin{align}
    \partial_{\theta_i} S(\eta \, || \, \rhot) = \mbox{Tr}(\eta \, H_i) - \mbox{Tr}(\rhot \, H_i),
    \label{eq:gradient-rule}
\end{align}
which is essentially the difference in expectation values of the terms that make up the Hamiltonian on the data and model density matrices. The derivation of the gradient is provided in Appendix~\ref{app:derivative-cost}. Recent results from Coopmans and Benedetti~\cite{coopmans_sample_2024} have shown that minimization of the quantum relative entropy with stochastic gradient descent and the fully-visible model is a convex problem and can be achieved using at most a polynomial (in system size $n$) number of Gibbs state preparations. Since the training procedure involves measuring expectation values of a pre-determined set of operators from a Gibbs state, classical shadows can be used to reduce the costs further~\cite{coopmans_predicting_2023}.

\subsection{\label{sec:tpq} Thermal pure quantum states}

Exact training of BMs and QBMs requires preparation of the Gibbs state defined in Eq.~\eqref{eq:gibbs}. This task quickly becomes expensive in terms of memory, especially in the case of a non-commuting Hamiltonian, as it requires exact diagonalization. While exact diagonalization can be used for small system sizes, we resort to approximate methods in order to scale our numerical results. In this section, we describe the numerical methods we use in this work.

Thermal pure quantum (TPQ) states are pure states, specified by a statistical ensemble, that are able to estimate properties such as expectation values of mixed states~\cite{PhysRevLett.108.240401}. For the Gibbs ensemble, a TPQ state $\ket{\psi}$ that is drawn at random satisfies
\begin{equation}
    \mbox{Pr}[|\bra{\psi} O_i \ket{\psi} - \mbox{Tr}(\rho_\beta O_i)| \geq \epsilon] \leq C_\epsilon e^{-\alpha n},
\label{eq:tpq-condition}
\end{equation}
where $\{ O_i \}$ is a set of Hermitian operators, $\rho_\beta$ is the Gibbs state at inverse temperature $\beta$ as defined in Eq.~\eqref{eq:gibbs} and $n$ is the system size. $C_\epsilon$ and $\alpha$ are constants that are not relevant for our purposes; therefore, we refer the reader to Ref.~\cite{coopmans_predicting_2023} for more details.

Coopmans~et~al.~\cite{coopmans_predicting_2023} have shown that pure states generated by imaginary time evolution,
\begin{equation}
    \ket{\psi_\beta} = \frac{e^{-\beta H /2} U \ket{0}}{\sqrt{\bra{0}  U^\dagger e^{-\beta H}  U\ket{0}}},
\end{equation}
satisfy Eq.~\eqref{eq:tpq-condition} with $U$ a random unitary drawn from the $n$-qubit Clifford group ($\mathcal{C}l(2^n)$). This leads to the following ensemble average to yield,
\begin{align}
    &\mathbb{E}_{U \thicksim \, \mathcal{C}l(2^n)} [\bra{\psi_\beta} O \ket{\psi_\beta}] \simeq \nonumber\\
    & \simeq \mbox{Tr}(\rho_\beta O) + \mbox{Tr}(\rho_\beta^2)\left( \mbox{Tr}(\rho_\beta O) - \mbox{Tr}(\rho_{2\beta} O)\right),
\label{eq:tpq-expectation}
\end{align}
where $O$ is a Hermitian operator. The detailed derivation and error analysis can be found in Ref.~\cite{coopmans_predicting_2023}. Recall that, in Eq.~\eqref{eq:gradient-rule}, we have shown that a fully-visible QBM can be trained by computing $\mbox{Tr} (\rhot H_i)$, where $H_i$ are the Pauli strings that form the Hamiltonian. Eq.~\eqref{eq:tpq-expectation} shows that this trace can be approximated by using TPQ states. This way, a QBM can be trained without the need to prepare the Gibbs state explicitly. Although this method was mainly developed to reduce quantum computational resources, it can also reduce the computational cost of simulations on a classical computer. 

Although TPQ states provide a cheaper computation to predict expectation values of Gibbs states compared to exact diagonalization methods, they do have limitations. One of the limitations we bring attention to is the finite errors of the model. For $\Tr(\rhot^2)<1$ the term proportional to the purity in Eq.~\eqref{eq:tpq-expectation} vanishes rapidly as $n\rightarrow\infty$ but it remains finite for pure states at any $n$~\cite{coopmans_predicting_2023}. Recall that in Eq.~\eqref{eq:qbm-target}, we have chosen the target state as a pure state for QBMs. This means that during the training of QBM, the system will get closer to a pure state (pure only when perfectly trained, mixed otherwise). Therefore, closer to convergence of the model, the TPQ states method will always yield systematic finite-size errors. In this work, we use the TPQ states method as an alternative to exact diagonalization to train QBMs.

Preparation of TPQ states on a classical computer also requires using a diagonalization method. Moreover, recall that the training procedure only requires estimating the expectation values over the Gibbs state. Although we can use TPQ states to train the model, they do not give access to the samples from the model. For this reason, in order to obtain the probability distribution of the model $\qt$, we need to diagonalize $\rhot$ or our estimate of it. We resort to the Lanczos diagonalization method~\cite{Lanczos:1950zz} in certain cases as a cheaper alternative compared to exact diagonalization.

The Lanczos method is an iterative approach, which allows diagonalization of matrix $A$ over the $D+1$ dimensional Krylov space $\mathcal{K}^D (\ket{v_i} = \text{span} \{  \ket{v_i}, A\ket{v_i}, A^2\ket{v_i}, \cdots, A^{D}\ket{v_i}   \}$, where $\ket{v_i}$ is the vector at step $i$ of the Lanczos iteration. The Lanczos method uses Krylov subspaces, allowing a cheaper but approximate diagonalization. The accuracy of the diagonalization depends on the choice of $D$, the Krylov dimension. 

\subsection{\label{sec:mut-info} Mutual information}

In Section~\ref{sec:trainining}, we made use of concepts from information theory, such as von Neumann entropy and quantum relative entropy. In the remainder of the work, we use conditional mutual information to reason about information spread in BMs and QBMs. 

Let us begin by defining the mutual information of a quantum system. Consider $\rho$ that describes the density matrix of a quantum system, which can be split into two subsystems that are denoted with $A$ and $B$. The bipartition is obtained via the partial trace, such that $\rho_A = \mbox{Tr}_B(\rho)$ and $\rho_B = \mbox{Tr}_A(\rho)$. Then, the mutual information of $\rho$ over the subsystems $A$ and $B$ is given as,
\begin{equation}
    \mathcal{I}(A:B) = S(A) + S(B) - S(AB),
    \label{eq:mut-info}
\end{equation}
where $S(A)$ and $S(B)$ are the von Neumann entropies of subsystems $\rho_A$ and $\rho_B$ and $S(AB)$ is the von Neumann entropy of the full system $\rho$. 

Conditional mutual information (CMI) describes the mutual information of two subsystems conditioned on another one~\cite{kuwahara_clustering_2024}. Let us consider $A$, $B$ and $C$, which are the subsystems of $\rho$. Then, CMI of subsystems $A$ and $C$ conditioned on $B$ is given as,
\begin{equation}
    \mathcal{I}(A:C|B) = S(AB) + S(BC) - S(ABC) - S(B),
    \label{eq:cmi}
\end{equation}
or equivalently, using Eq.~\eqref{eq:mut-info}, it can be written as,
\begin{equation}
    \mathcal{I}(A:C|B) = \mathcal{I}(A:BC) - \mathcal{I}(A:B).
\end{equation}

Recent work by Kuwahara has proposed that the CMI of quantum systems vanish exponentially in distance and the correlation length grows polynomially in inverse temperature~\cite{kuwahara_clustering_2024}. 

\section{\label{sec:results} Numerical results}

This section presents numerical results analyzing the capabilities of fully-visible QBMs to learn various target distributions. Furthermore, we provide numerical evidence for instances where the tested fully-visible QBMs achieve lower KL divergence compared to the tested fully-visible BMs, indicating a better model performance of the QBM. As target distributions, we choose randomly generated Boltzmann distributions with up to third-order interactions and a real-world dataset related to particle jet events from high energy physics.

We compute the density matrices of BMs using exact diagonalization for all demonstrations. We use the TPQ states~\cite{coopmans_predicting_2023} and Lanczos methods~\cite{Lanczos:1950zz} to approximate QBMs when stated; otherwise, we use exact diagonalization. We set the inverse temperature $\beta=1$. All models are trained using the \textsc{AMSGrad} optimizer ($\mbox{lr}=0.1, \beta_1=0.9, \beta_2=0.99$)~\cite{amsgrad} for 1000 steps. We initialize all models with zero initial parameters. The choice of initial point does not impact the results significantly due to the convex loss landscape we consider~\cite{coopmans_sample_2024}. Since our goal is to provide a comparison of BM and QBM models on equal footing rather than to present a fully-trained model, we do not perform hyperparameter optimization. We note that the default hyperparameters are sufficient for all models to converge. The data and the code to reproduce the plots can be found in Ref.~\cite{open-data}.

\subsection{\label{sec:exp-boltzmann} Learning Boltzmann distributions}

\begin{figure}[!t]
    \centering
    \includegraphics[width=\linewidth]{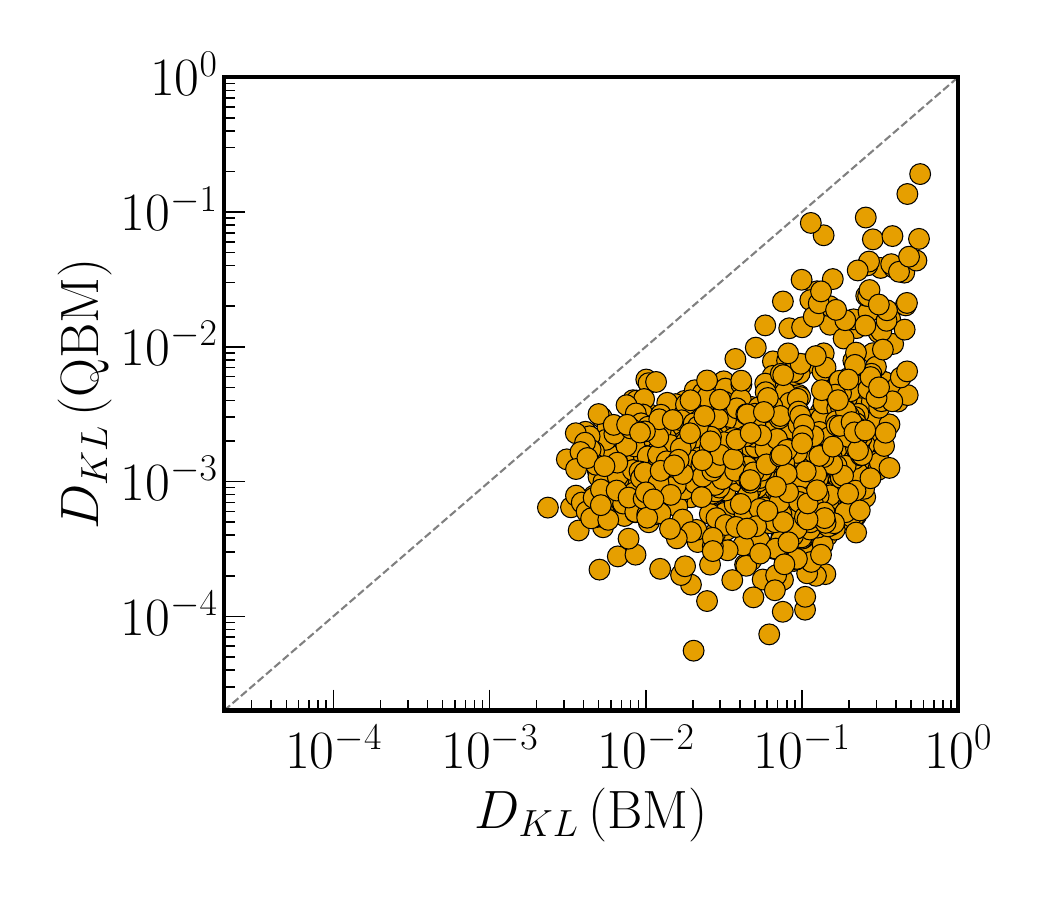}
    \caption{\textbf{BM vs. QBM (\textit{generic}) $\KL$ values after being trained on three-body Boltzmann distributions.} Four qubit, all-to-all connected BM and QBM  models are trained using exact methods and $\KL$ is measured after training. 1000 target distributions are used, which are sampled according to Eq.~\eqref{eq:boltzmann-dist} with $\parallel w^{(1)}\parallel_1 = 1, \, \parallel w^{(2)}\parallel_1 = 5, \, \parallel w^{(3)}\parallel_1 = 5$ interaction strength.}
    \label{fig:kl-boltzmann-1-5-5}
\end{figure}

This section presents a numerical analysis of the difference in learning capacity of fully-visible QBMs and BMs using randomly generated Boltzmann distributions with different underlying graphs as targets. We consider the Boltzmann distribution defined by
\begin{equation}
p(s) = \frac{e^{-\beta E(s)}}{\sum_s e^{-\beta E(s)}}, 
\end{equation}
where the energy function is given as
\begin{equation}
E(s) =  \sum_{i=0}^{n-1} w^{(1)}_i s_i + \sum_{\substack{i,j=0 \\ i \neq j}}^{n-1} w^{(2)}_{i,j} s_i s_j + \sum_{\substack{i,j,k=0 \\ i \neq j \neq k}}^{n-1} w^{(3)}_{i,j,k} s_i s_j s_k,
\label{eq:boltzmann-dist}
\end{equation}
where $s = s_0 s_1 \cdots s_{n-1}$ and $s_i \in \{-1, +1\}$ for all $i$. In Eq.~\eqref{eq:boltzmann-dist}, we provide the energy function with up to three-body interactions for simplicity. This definition can be restricted or extended to other types of interactions, from single-body to $n$-body interactions. Each of these terms is parametrized with $w^{(k)}$, which has $\binom{n}{k}$ unique entries for an all-to-all connected graph. We normalize the parameters such that the contribution of each $k$-body interaction can be controlled. We choose $\parallel w^{(1)}\parallel_1 = 1, \, \parallel w^{(2)}\parallel_1 = 5, \, \parallel w^{(3)}\parallel_1 = 5$ on the all-to-all connected graph with four vertices ($n=4$). This choice allows the two-body and three-body interactions to dominate the spectrum. We sample 1000 sets of parameters from the uniform distribution between $[-1, 1]$, not to favor any configuration. Using the set of random parameters, we generate 1000 unique probability distributions to be used as target distributions. We learn the generated distributions using a four qubit, all-to-all connected BM and a four qubit, all-to-all connected QBM equipped with the \textit{generic} Hamiltonian, which includes all combinations of one- and two-body Pauli matrices (cf. Tab.~\ref{tab:hamiltonian-defs}). We report the final $\KL$ in Fig.~\ref{fig:kl-boltzmann-1-5-5}.

Recall that both the BM and QBM are constructed with up to two-body interactions, while the target distribution is constructed with up to three-body interactions. Results presented in Fig.~\ref{fig:kl-boltzmann-1-5-5} illustrate that the QBM outperforms the BM in the task of learning the target distributions. This is attributed to the fact that all distributions we consider here have non-zero third-order correlations. A fully-visible BM with two-body interactions is naturally expected to learn distributions with only up to two-body interactions. However, QBMs are not limited in the same way and, as we demonstrate, can learn higher-order distributions much better than BMs.

One might think that QBMs can outperform BMs in this task simply due to the fact that they have more parameters. Here, we emphasize that although QBMs have more parameters, their representation power comes from the non-commuting terms they contain. BMs can alternatively be made more expressive by considering higher-order interactions, but this would increase the computational costs significantly. In order to make the learning capability separation between BMs and QBMs more clear, we propose a second learning task.

We choose a setting where both models and target distributions are built with up to two-body interactions, while we restrict the connectivity of the models as well as the target. For this purpose, we define a probability distribution generated by a next-nearest-neighbor Boltzmann distribution on a chain with eight sites. With the next-nearest neighbor connections, the target distribution becomes two-dimensional. We define the dimension of a distribution or a model as the dimension of their lattice; e.g., a line is one-dimensional, while a grid is two-dimensional. It is important to note that the dimension of the model is different than the order of interactions it contains. For example, a model with one- and two-body interactions can be constructed to be one-dimensional (on a chain) or $n-1$-dimensional (all-to-all connected). A visualization of the connectivity is presented in Fig.~\ref{fig:8q-next-nn} of the appendix. The energy function of the target distribution with $n$ sites and open boundary conditions is given as
\begin{equation}
E(s) =  \sum_{i=0}^{n-1} s_i + \sum_{i=0}^{n-2} s_i s_{i+1} + 0.5 \times \sum_{i=0}^{n-3} s_i s_{i+2}.
\label{eq:next-nn-energy}
\end{equation}

We train two types of BMs and QBMs on the eight site next-nearest-neighbor target distribution. BM uses the \textit{ising} and QBM uses the \textit{generic} Hamiltonian defined in Table~\ref{tab:hamiltonian-defs}. The first type is the all-to-all connected model. In this case, both the BM and QBM can perfectly learn ($\KL=0$) and represent the next-nearest-neighbor target distribution. Having all-to-all connectivity allows both models to represent high-dimensional target distributions. In the second type, we restrict the connectivity of both models to nearest-neighbor (NN) with open boundary conditions. This way, both models are restricted to one dimension, while the target has two dimensions. In this case, the QBM ($\KL=0.15$) approximates the target distribution much better than the BM ($\KL=0.4$).

\begin{figure}[!t]
    \centering
    \includegraphics[width=\linewidth]{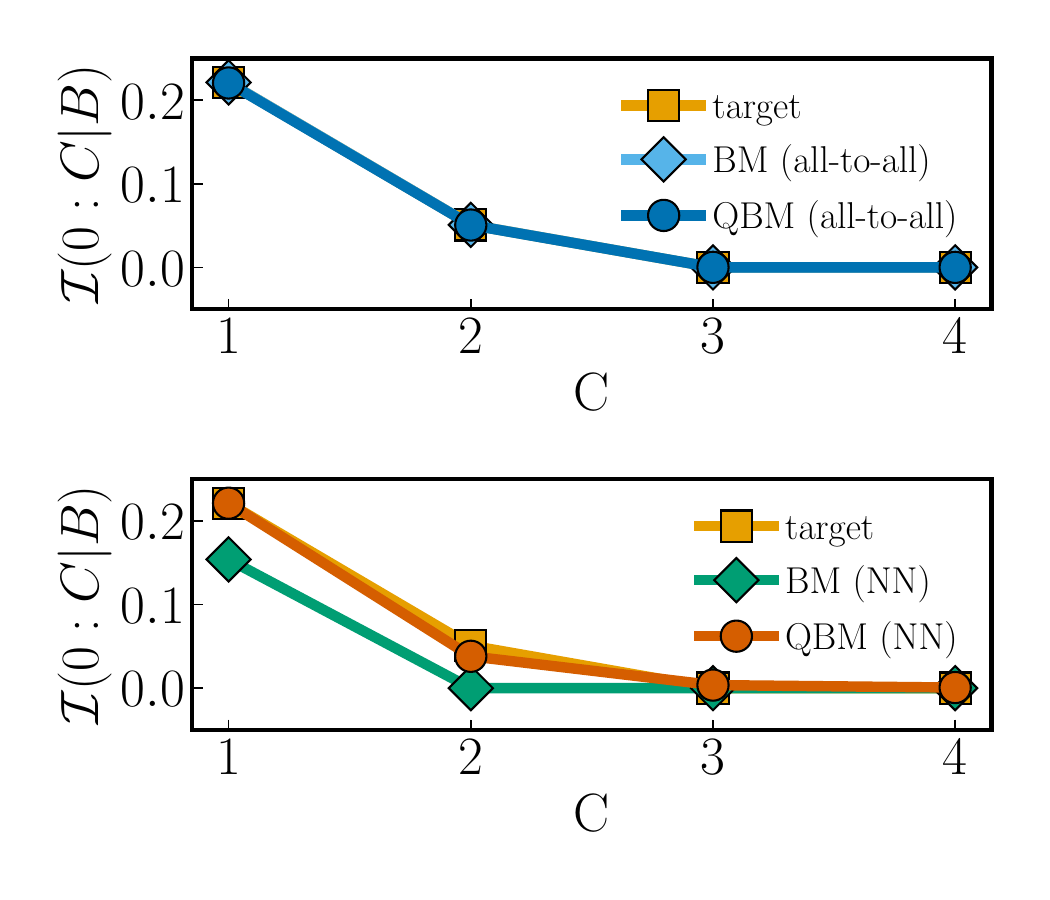}
    \caption{\textbf{Conditional mutual information (CMI) measured on the distributions of the models trained on the next-nearest-neighbor distribution.} The target distribution is the next-nearest-neighbor distribution defined in Eq.~\eqref{eq:next-nn-energy}. (\textbf{top}) Results for models with all-to-all connectivity (all data coincide). (\textbf{bottom}) Results for models with nearest-neighbor (NN) connectivity. More details on model connectivity are provided in Fig.~\ref{fig:connectivity} of the appendix.}
    \label{fig:1d-boltzmann}
\end{figure}

The difference between BM and QBM becomes more apparent when we observe the CMI (recall Eq.~\eqref{eq:cmi}) produced by the probability distribution of the trained models. Let us choose the target qubit index 0 and obtain the CMI with respect to the nearest four qubits $C \in \{1,2,3,4\}$. We plot the CMI of the four cases in Fig.~\ref{fig:1d-boltzmann} along with the target distribution. As the target distribution is generated by a next-nearest-neighbor model, the CMI is highest on index one ($C=1$) and steadily decays with the distance. It is clear that the all-to-all connected models can represent the mutual information spread of the target. In the NN-connected case, the conditional mutual information for the NN-connected BM vanishes for index $C\geq 2$ as it does not have the necessary connections. In contrast, the QBM has non-zero conditional mutual information on index $C=2$, although it does not have the connection between sites zero and two. This shows that fully-visible QBMs can learn distributions that have higher dimensions than themselves.

\subsection{\label{sec:jet} Learning particle jet events}

This section presents numerical evidence that the results from randomly generated target distributions in the preceding subsection can be extended to real-world examples from particle physics.

We use the JetNet dataset~\cite{kansal_2022_6975118} to produce probability distributions to compare the performance of BMs and QBMs. The JetNet dataset consists of collections of particle jet event data that are simulated based on the Standard Model. We refer the reader to Ref.~\cite{kansal_particle_2021} for details of the simulations. Particle jets are clusters of particles that are often observed at particle collider experiments. Particles of each jet lead to high-dimensional and highly correlated distributions. Being able to simulate and sample from these distributions is of high interest to the high energy physics community~\cite{cms_collaboration_search_2020}.

In this work, we focus on the absolute relative transversal momentum ($\pt$) of jets that originate from W bosons. It is important to note that different types of jets lead to different distributions; however, the type of the jet is not relevant for this study. We choose the $m$ highest $\pt$ particles while forming the target distributions. The continuous values of $\pt$ for each particle are used to construct a multi-dimensional histogram with $k$ equally-split bins bounded by the minimum and maximum values observed in the dataset. By normalizing the histogram, we construct an estimator for the joint probability distribution of $m$ particles. This way, we represent one feature of a jet that consists of $m$ particles with $m \times \log_2(\nbins)$ bits. This allows us to use a small system to represent the same problem, but with low precision. Using $\sim 100,000$ jet events from JetNet, we construct train, test and validation distributions (with $0.7/0.15/0.15$, train/test/validation ratio). The train distributions are used only during the training stage and the test distribution is used to assess the learning performance after training. We provide a visualization of the dataset for the $m=4$ particle and $\nbins=16$ case in Fig.~\ref{fig:jetnet-distribution}. In Fig.~\ref{fig:mutinfo-target} of the appendix, we present mutual information, measured on the dataset, in order to demonstrate its highly-correlated nature.

\begin{figure}[!t]
    \centering
    \includegraphics[width=\linewidth]{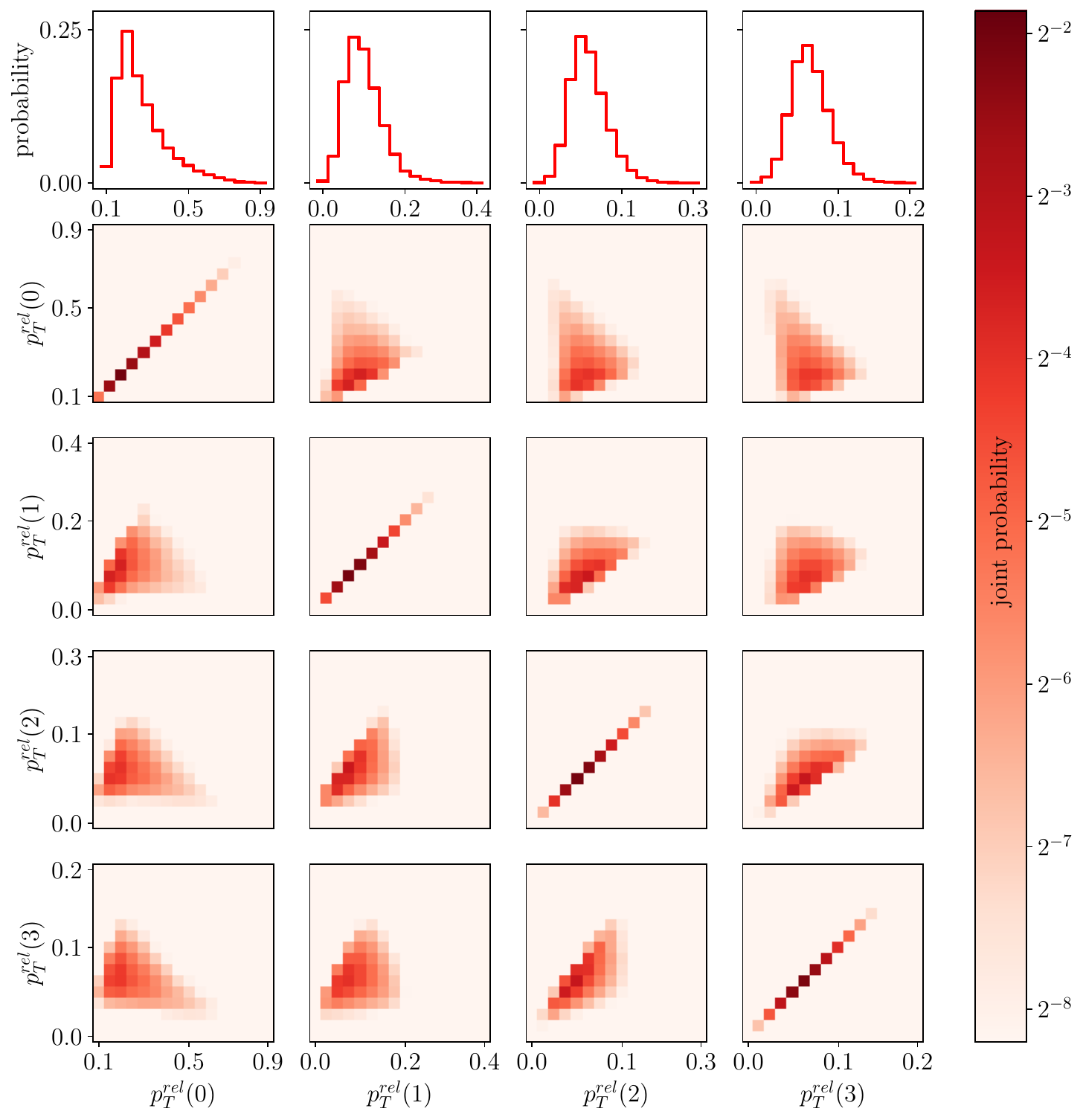}
    \caption{\textbf{One and two dimensional projections of the $m=4$ particle joint probability distribution.} Top row corresponds to the marginal distribution of each particle. Other rows (except the diagonal) are the two particle joint distributions. All distributions are obtained by normalizing the corresponding histogram. Histograms contain $\nbins = 16$ per particle. The minimum and maximum values are varied for each particle to avoid empty bins as much as possible.}
    \label{fig:jetnet-distribution}
\end{figure}

\subsubsection{Boltzmann machines vs. quantum Boltzmann machines}

In this section, we compare the BM and QBM learning capabilities with various settings on the particle jet event generation problem. 

We report all the results using exact methods for BMs and the TPQ state method for QBMs. Since BMs have diagonal Gibbs states, we are able to use exact methods for the system sizes we are considering. However, a classical computer simulation of QBMs requires significantly more computational power than BMs due to the non-diagonal Gibbs states. 

To show the accuracy of the TPQ states method, we begin by comparing it to the exact diagonalization on a ten qubit problem. We choose as target the discretized $\pt$ distribution with $m=2$ particles and $n_\text{bins}=32$ ($n=10$ qubits). The QBM uses the \textit{generic} Hamiltonian (cf. Tab.~\ref{tab:hamiltonian-defs}) with all-to-all connectivity. QBM training uses 100 TPQ states and the Lanczos method with $D=20$, where $D$ is the Krylov dimension.

\begin{figure}[!t]
    \centering
    \includegraphics[width=\linewidth]{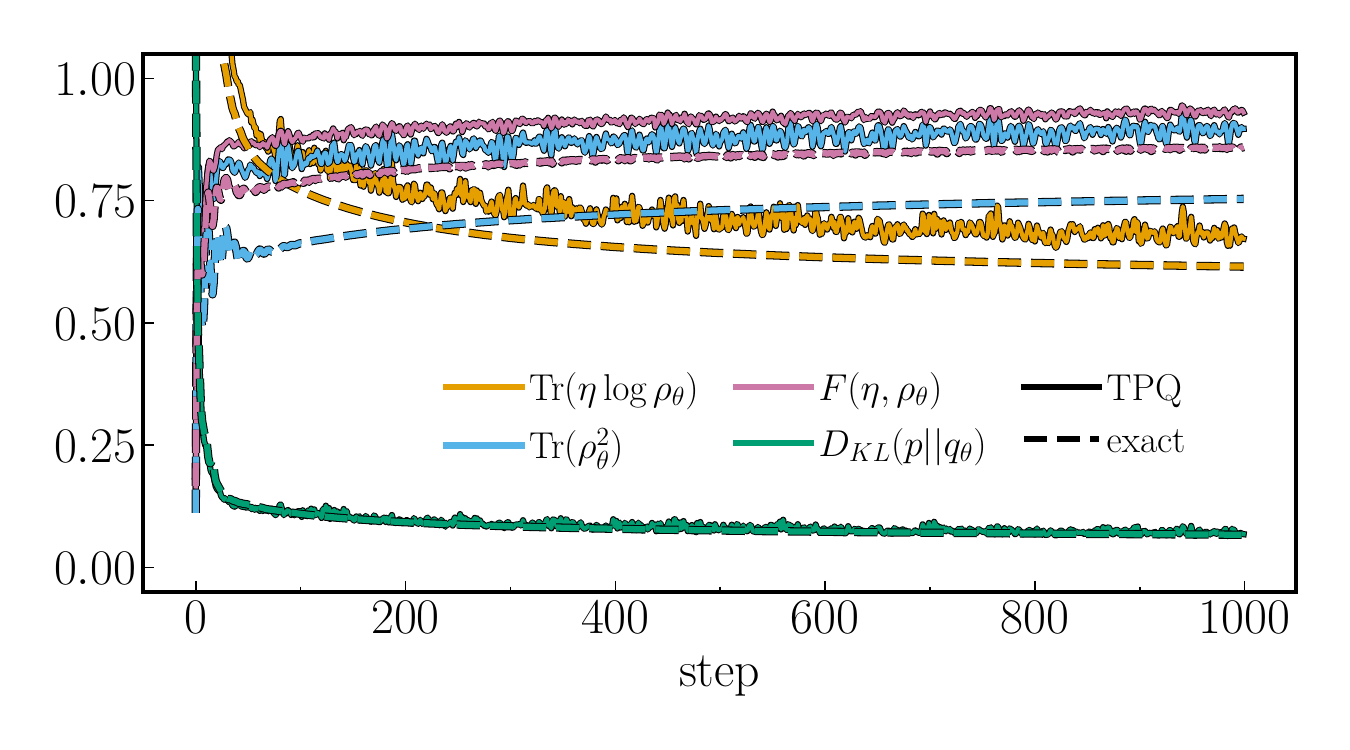}
    \caption{\textbf{QBM training curve comparison with exact vs. TPQ states methods.} The target distribution is the $\pt$ distribution of the $m=2$ particle case with $\nbins = 32$ ($n=10$ qubits). QBM has all-to-all connectivity and uses the \textit{generic} Hamiltonian. The TPQ states training method uses 100 TPQ states and the Lanczos method with $D=20$.}
    \label{fig:tpq-hist}
\end{figure}

We present several metrics to assess the training performance using TPQ and exact diagonalization methods during training in Fig.~\ref{fig:tpq-hist}. We observe that the training performance with TPQ states can match the training performance with exact diagonalization. The negative quantum log-likelihood ($\mbox{Tr}(\eta \, \mbox{log}\, \rhot)$) decreases monotonically and converges to the value of the negative von Neumann entropy of the target state ($\mbox{Tr}(\eta \, \mbox{log}\, \eta)$). Recall from Eq.~\eqref{eq:quantum-rel-entropy} that this means the cost function, quantum relative entropy, reaches zero. The KL divergence from the model distribution $\qt$ to target $p$ ($\KL(p || \qt)$) converges to zero. The fidelity between model and target density matrices ($F(\eta, \rhot)$), as well as the purity of the model state ($\mbox{Tr} (\rhot^2)$) both approach one. The accuracy of the results is highly dependent on the number of TPQ states and the Krylov dimension chosen. We have decided that the chosen values are sufficient for the range of system sizes we consider, after an empirical assessment, which can be found in Appendix~\ref{app:tpq-sweep}.

\begin{figure}[!t]
    \centering
    \includegraphics[width=\linewidth]{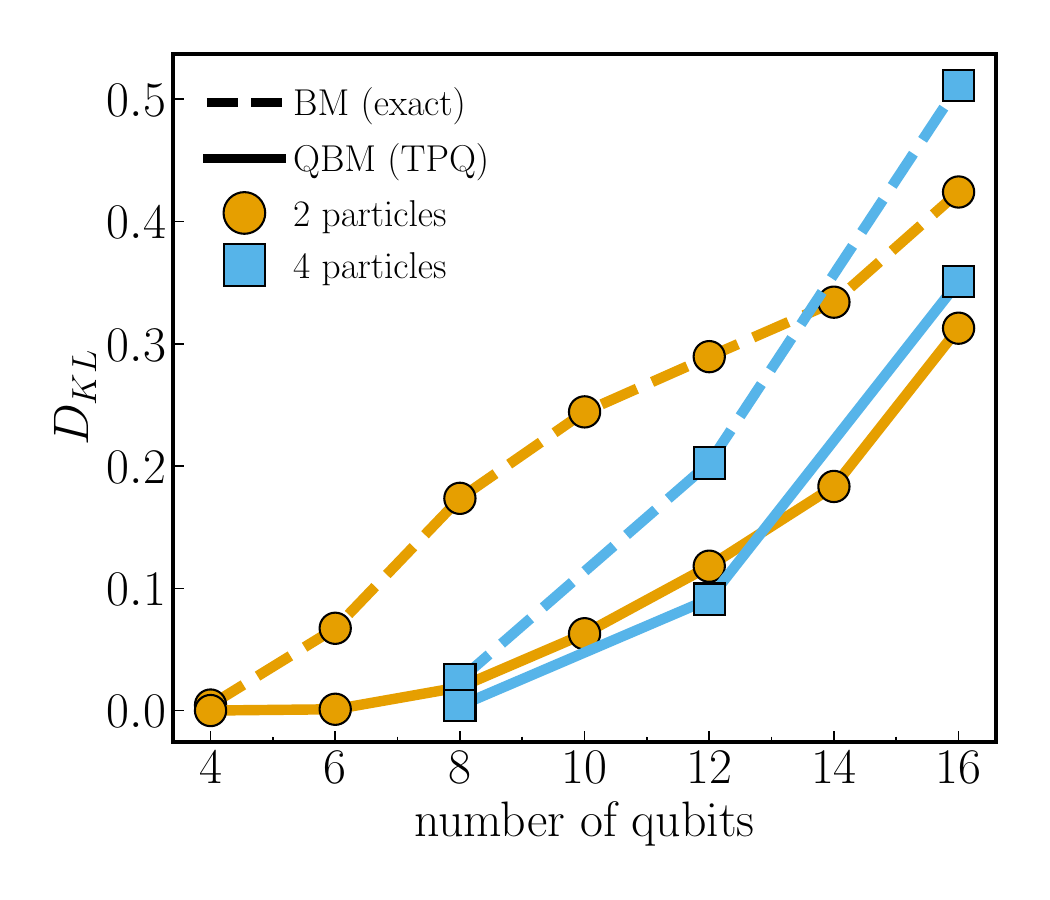}
    \caption{\textbf{Best $\KL$ measured after training on datasets obtained for $m=2$ particles (orange circles) and $m=4$ particles (blue squares) with different numbers of bins using BM (dashed) and QBM (solid).} BM is trained using the exact diagonalization, while QBM is trained using the TPQ states method. The QBM model uses the \textit{generic} Hamiltonian. It is trained using 100 TPQ states and the expectation values are estimated using the Lanczos method with $D=20$. All models are all-to-all connected.}
    \label{fig:kl-jetnet}
\end{figure}

After establishing the accuracy of the TPQ states method to train QBMs, we provide results for various target states in Fig.~\ref{fig:kl-jetnet} using all-to-all connected BMs and QBMs of various sizes. We choose the target states with $m=2$ and $m=4$ particles and a varying number of bins. We observe that in all of the cases, QBMs can reach lower $\KL$ compared to BMs, which are trained using exact diagonalization. This result provides evidence for the utility of QBMs in outperforming BMs in (small-scale) real-world problems. 

Following these results, we consider BMs and QBMs with different connectivity. In Section~\ref{sec:exp-boltzmann}, we have established that QBMs are capable of learning higher dimensional distributions. This time, we test this hypothesis on the particle physics data. The dataset that we are considering is a good test bed, as previous findings have shown improved results with all-to-all connected classical GNNs~\cite{kansal_particle_2021}. We illustrate the correlations by measuring the mutual information on the target distribution in Fig.~\ref{fig:mutinfo-target} of the appendix as mentioned before.

We choose two connectivity settings for BMs and QBMs. The first one is the all-to-all connectivity. The second one is a connectivity that we denote as nearest-neighbor-particle (NN-particle). The NN-particle setting connects all units that belong to the same particle to each other, while the units of different particles are connected in a nearest-neighbor fashion (ordered by their respective $\pt$). We present an illustration of all-to-all and NN-particle settings in Fig.~\ref{fig:conn-all-to-all} and Fig.~\ref{fig:conn-nn-particle} of the appendix.

\begin{figure}[!t]
    \centering
    \includegraphics[width=\linewidth]{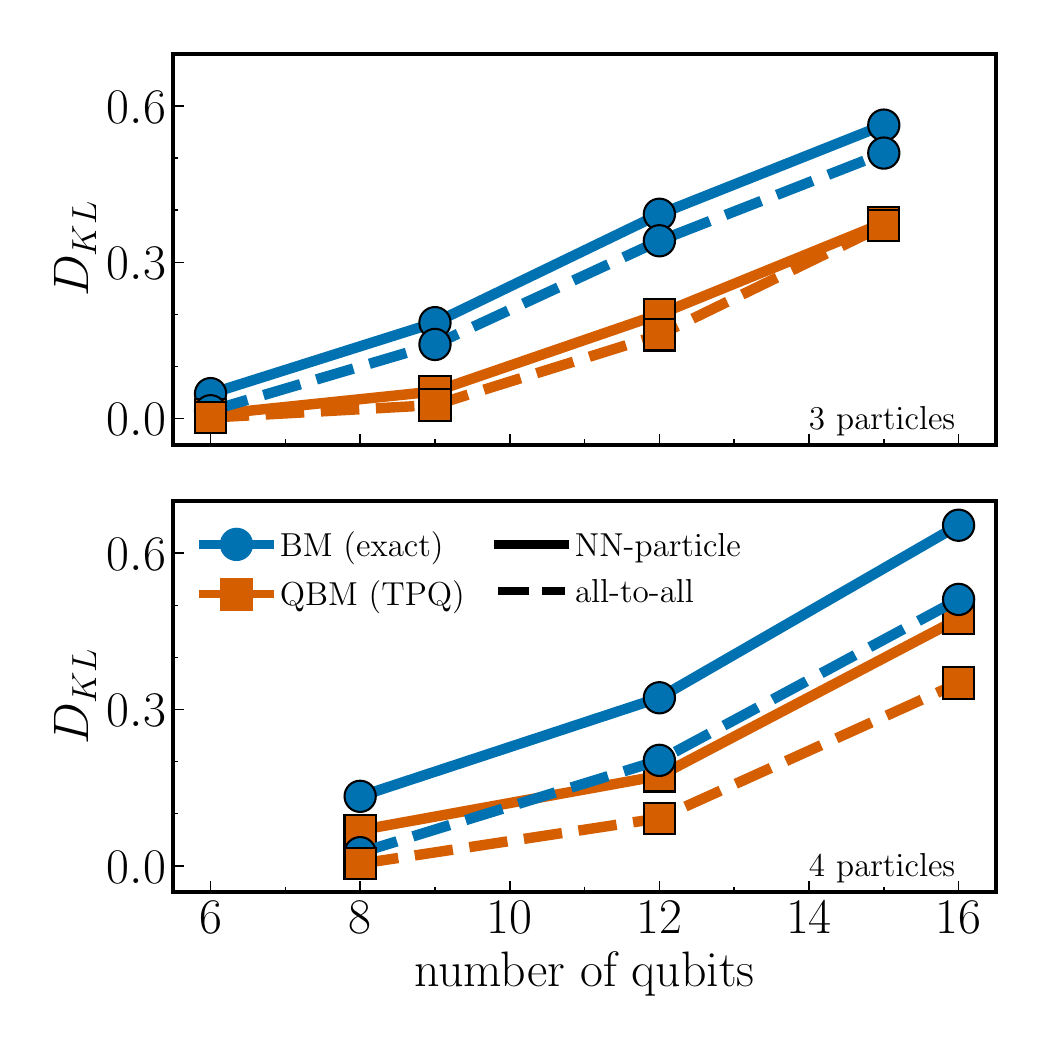}
    \caption{\textbf{Connectivity comparison of BM and QBM.} BM and QBM models are trained on $m=3$ and $m=4$ particle target distributions using all-to-all and nearest-neighbor-particle (NN-particle) connectivity. More details regarding connectivity can be found in Fig.~\ref{fig:conn-nn-particle} of the appendix. The BM is trained using exact diagonalization, while the QBM is trained using the TPQ states method. The QBM uses the \textit{generic} Hamiltonian and is trained using 100 TPQ states with the Lanczos diagonalization method with $D=20$.}
    \label{fig:compare-conn}
\end{figure}

We present training results for the BM and QBM using the two types of connectivity on three particle and four particle datasets in Fig.~\ref{fig:compare-conn}. We observe a similar pattern as before, in which the QBM represents the distributions better than the BM. However, what is more striking is that the QBM with limited connectivity (NN-particle) can still represent the distributions at least as well as the all-to-all connected BM. 

Note that in all instances, the $\KL$ values increase with increasing number of qubits. This does not indicate decreasing performance but is an effect of the change in system size. $\KL$ values should only be compared against the same target state with the same system size; otherwise, the comparison may be misleading, as it is an unbounded metric. 

\subsubsection{Improving quantum Boltzmann machine performance}

So far, we have considered QBMs only with the $\textit{generic}$ Hamiltonian. This is because this choice is the most expressive one that leads to the best representation capability among other choices listed in Table~\ref{tab:hamiltonian-defs}. Previous works have used \textit{tfim} and \textit{spin-glass} Hamiltonians due to their connections to quantum many-body physics~\cite{amin_quantum_2018, kappen_learning_2020}. However, our results show that the $\textit{generic}$ Hamiltonian that contains all possible weight one and two Pauli strings outperforms them in terms of the representation capability. This does not appear to be only due to having more parameters but also due to adding additional degrees of freedom to the model through the non-commuting terms. We have shown the effect of non-commuting terms more explicitly in Section~\ref{sec:exp-boltzmann}, when we discussed the CMI of a one-dimensional BM and QBM. We present a detailed comparison of the choice of Hamiltonian in Appendix~\ref{app:ham-comp}. 

A natural question arises: after establishing the representation capability of the $\textit{generic}$ Hamiltonian, is there a Hamiltonian that has the same representation capability with fewer terms or parameters? We answer this in the affirmative. Recall that we chose the phase of the embedding in Eq.~\eqref{eq:qbm-target} to be zero. This means that the model $\rhot$ will only consist of real entries. Therefore, the terms of the Hamiltonian that contain imaginary values will always have zero expectation values. These terms are the ones that contain an odd number of Pauli-Y operators. For this reason all the terms that contain an odd number of Pauli-Y operators can be omitted from the Hamiltonian definition. As a result, the \textit{generic} Hamiltonian can be reduced to the \textit{generic-real} Hamiltonian by excluding the terms such as $\{ Y, XY, YX, ZY, YZ \}$, while keeping the same representation capability. A similar reduction can be applied to the \textit{spin-glass} Hamiltonian used by Kappen~\cite{kappen_learning_2020}. This observation reduces the number of parameters and may reduce resource requirements for implementation of the Gibbs state on quantum hardware.

It is possible to approach this question from a different angle. Since each Hamiltonian term $H_i$ is parametrized with a scalar parameter, the magnitude of the parameter is a measure of its significance in the total Hamiltonian. We prune the terms of a trained QBM with \textit{generic-real} Hamiltonian at various thresholds and report the $\KL$ in Table~\ref{tab:pruning}. This is equivalent to assigning a value of zero to parameters with magnitudes below the threshold. We observe that approximately $10\%$ of the terms can be further removed without leading to a significant loss in quality of the output distribution. Such a simple pruning may help reduce the costs of implementing the Gibbs state during and after training. It is also important to note here that the terms to be pruned are problem-dependent and are unknown prior to training. An alternative strategy can be to track the size of the gradients and prune the terms that have small gradients after a few training steps, or use ideas from L1 regularization~\cite{tibshiraniRegressionShrinkageSelection1996} which has been applied to reduce parameters in variational quantum algorithms~\cite{duffieldBayesianLearningParameterised2023}. We leave this as future work.

\begin{table}[!t]
    \centering
    \caption{\textbf{Effect of weight pruning to model performance after training.} We train a QBM with the \textit{generic-real} Hamiltonian using TPQ states on the $m=2$ particle and $\nbins = 32$ data ($n=10$ qubits). All $\KL$ values are evaluated using exact methods to isolate errors from the TPQ states method.}
    \begin{tabularx}{\linewidth}{X|X|X}
         Threshold & $\KL$ & Terms removed $[\%]$\\
         \hline
         \hline
         $0.0$ & $8.6 \times 10^{-2}$ & $0 \%$\\
         \hline
         $1.0 \times 10^{-3}$ & $8.6 \times 10^{-2}$ & $0.4 \%$\\
         \hline
         $5.0 \times 10^{-3}$ & $8.6 \times 10^{-2}$ & $0.7 \%$\\
         \hline
         $1.0 \times 10^{-2}$ & $8.5 \times 10^{-2}$ & $2.0 \%$\\
         \hline
         $5.0 \times 10^{-2}$ & $8.6 \times 10^{-2}$ & $6.1 \%$\\
         \hline
         $1.0 \times 10^{-1}$ & $8.7 \times 10^{-2}$ & $11.0 \%$\\
         \hline
         $5.0 \times 10^{-1}$ & $5.2 \times 10^{-1}$ & $49.4 \%$\\
         \hline
         $1.0$ & $2.1$ & $68.2 \%$\\
         \hline
    \end{tabularx}

    \label{tab:pruning}
\end{table}

Another important factor to consider regarding the parameters is the inverse temperature $\beta$. So far, we have assumed $\beta=1$ and considered that it is a global term acting on all of the Hamiltonian parameters $\mathbf{\theta}$, such that $\mathbf{\theta}$ is unbounded. Since $\mathbf{\theta}$ is unbounded, this induces an effective inverse temperature $\Tilde{\beta}$ that we define as
\begin{equation}
    \Tilde{\beta} = \mbox{max} \left( | \mathbf{\theta}| \right),
\end{equation}
where $|\cdot|$ denotes an element-wise absolute value. As the last row in Table~\ref{tab:pruning} shows, there exist parameters with an absolute value larger than one. This indicates that the effective inverse temperature of the model rises during training, which starts with all parameters initialized to zero. This is expected since the target state of a QBM is encoded as a pure state, as in Eq.~\eqref{eq:qbm-target}. This fact naturally brings up a discussion on the value of $\Tilde{\beta}$ and whether it can be considered as a hyperparameter of the model. In Appendix~\ref{app:sec:temperature}, we provide extended numerical results showing that higher values of $\Tilde{\beta}$ provide only marginal improvement, while lower values only make the results worse. This implies that the training procedure finds an optimal effective inverse temperature. Since the cost of Gibbs state preparation increases with the inverse temperature~\cite{chenQuantumThermalState2023}, it is favorable to find the lowest inverse temperature that produces the same result.

\section{\label{sec:conclusion} Discussion}

In recent years, researchers have proposed using quantum systems to build variations of BM models. This includes fully-visible and restricted QBMs. Restricted QBMs have been shown to be difficult to train due to the non-commuting nature of the system Hamiltonian~\cite{amin_quantum_2018, wiebe_generative_2019, kieferova_tomography_2017}. On the other hand, recent results have shown that fully-visible QBMs can be trained with a number of Gibbs states polynomial in the system size~\cite{coopmans_sample_2024}. This puts fully-visible QBMs in a sweet spot with respect to fully-visible BMs, which have limited learning capacity and restricted QBMs, which are difficult to train. In this work, we aim to understand the extent of the learning capabilities of fully-visible QBMs. We demonstrate that fully-visible QBMs offer advantages over fully-visible BMs, particularly in terms of learning capacity. Specifically, we numerically show that fully-visible QBMs can capture complex distributions that involve higher-order interactions and increased connectivity. \looseness-1

While these findings are based on constructed examples where BMs struggle by design, we also demonstrate their relevance to real-world datasets. A compelling example is particle jet event generation, where data originates from highly correlated particles, resulting in a complex, high-dimensional underlying distribution. This is a setting where fully-visible BMs often fail, yet QBMs excel due to their enhanced expressivity. 

We also show that Hamiltonians previously used for QBMs limit their performance, and adopting more general Hamiltonians significantly improves learning capabilities. 

The results we present using TPQ states, as described in Section~\ref{sec:tpq}, do include some systematic errors. This suggests the potential for even better QBM performance in future studies, implying that the learning capabilities we report here may be far from the theoretical maximum. However, since we only present numerical results for small system sizes, these models must be evaluated at a larger scale to determine if our findings are applicable to a broader spectrum of problems.

Our results contribute to a growing momentum in exploratory QML towards novel models beyond parametrized quantum circuits or kernel methods. Recent studies, including ours, suggest that non-unitary approaches~\cite{heredge2024}, such as QBMs, could play a crucial role in advancing QML~\cite{cerezo2023does}.

Despite the promising results on small system sizes, QBMs require fault-tolerant quantum devices with hundreds of logical qubits to solve practically relevant problems. Achieving this will demand collaborative efforts from both experimentalists and theorists. The recent surge in quantum algorithms for Gibbs state preparation (see e.g., \cite{PhysRevLett.103.220502, temme_quantum_2011, 10.5555/3179483.3179486, holmesQuantumAlgorithmsFluctuation2022,wocjanSzegedyWalkUnitaries2021,rallThermalStatePreparation2023,zhangDissipativeQuantumGibbs2023,chenQuantumThermalState2023, chenEfficientExactNoncommutative2023,dingSimulatingOpenQuantum2024,gilyenQuantumGeneralizationsGlauber2024,dingEfficientQuantumGibbs2024,chenRandomizedMethodSimulating2024}) together with hardware advances could push future studies of the QBMs to larger and larger sizes.

In future work, a more comprehensive comparison of fully visible QBMs with state-of-the-art GNN methods would be valuable. Additionally, in this work, we considered real-world applications only in particle physics. It would be worthwhile to investigate datasets from other fields where QBMs can demonstrate potential benefits.\\

\begin{acknowledgments}
We thank Marcello Benedetti and Ifan Williams for their feedback on an earlier version of this manuscript. C.T. is supported in part by the Helmholtz Association -``Innopool Project Variational Quantum Computer Simulations (VQCS)''. This work is supported with funds from the Ministry of Science, Research and Culture of the State of Brandenburg within the Centre for Quantum Technologies and Applications (CQTA). This work is funded within the framework of QUEST by the European Union’s Horizon Europe Framework Programme (HORIZON) under the ERA Chair scheme with grant agreement No.\ 101087126. This work was supported by the Netherlands Organisation for Scientific Research (NWO/OCW), as part of the Quantum Software Consortium programme (project number 024.003.037 / 3368).
\begin{figure}[!h]
    \centering
    \includegraphics[width=0.1\textwidth]{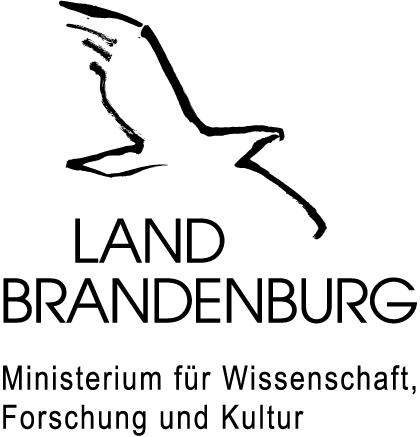}
\end{figure}
\end{acknowledgments}

\bibliography{main}

\onecolumngrid
\appendix

\section{Gradient derivation}

\subsection{Derivative of the matrix exponential and trace}

The derivative of the matrix exponential is defined as follows:
\begin{equation}
    \frac{d}{dt}e^{X(t)} = e^{X(t)}\frac{1-e^{-\text{ad}_X}}{\text{ad}_X} \frac{dX(t)}{dt},
\end{equation}
and $\text{ad}_X$ is given as $\text{ad}_X ( \cdot) = \left[ X, \cdot \right]$. Then, we can also write the following power series,
\begin{equation}
\frac{1-e^{-\text{ad}_X}}{\text{ad}_X} = \sum_{k=0}^{\infty} \frac{(-1)^k}{(k+1)!} \left( \mbox{ad}_X \right)^k.
\end{equation}

Last but not least, recall that the Hamiltonian has the following form,
\begin{equation}
    \Ht = \sum_i \theta_i H_i,
\end{equation}
where $\theta_i$ are real valued parameters and $H_i$ are Pauli strings. Then, we combine these results to obtain,
\begin{align} 
    \partial_{\theta_i} e^{-\Ht} &= e^{-\Ht} \sum_{k=0}^{\infty} \frac{(-1)^k}{(k+1)!} \left( \mbox{ad}_{(-\Ht)} \right)^k (-H_i) \nonumber \\
    &= e^{-\Ht} \left( -H_i - \frac{1}{2} \mbox{ad}_{(-\Ht)}(-H_i) + \frac{1}{6} \mbox{ad}_{(-\Ht)} \mbox{ad}_{(-\Ht)}(-H_i) + \cdots \right) \nonumber \\
    & = e^{-\Ht} \left( -H_i - \frac{1}{2} [\Ht, H_i] - \frac{1}{6} [\Ht, [\Ht, H_i]] + \cdots \right).
\end{align}

Here one can observe that only the leading term is sufficient in the case of a commuting Hamiltonian. For Hamiltonians with non-commuting terms, we need to compute the nested set of commutators. Fortunately, we can enjoy a nice property of the trace and the commutators to avoid computing all of the remaining terms when computing the trace of the derivative ($\mbox{Tr} (\partial_{\theta_i} e^{-\Ht} )$). 

Let us show this for the first commutator that appears for a trace of the form $\mbox{Tr} (e^{-\Ht} [\Ht, H_i] )$. Observe that we can write this as
\begin{align}
    \mbox{Tr} (e^{-\Ht} [\Ht, H_i] ) &= \mbox{Tr} (e^{-\Ht} \Ht H_i) - \mbox{Tr} (e^{-\Ht} H_i \Ht) \\
    &= \mbox{Tr} (\Ht e^{-\Ht} H_i) - \mbox{Tr} (e^{-\Ht} H_i \Ht) \label{app:eq:trace-1}\\
    &= \mbox{Tr} (e^{-\Ht} H_i \Ht) - \mbox{Tr} (e^{-\Ht} H_i \Ht) = 0. \label{app:eq:trace-2}
\end{align}

In Eq.~\eqref{app:eq:trace-1} we alternate $\Ht$ and $e^{-\Ht}$. This is possible as these terms commute with each other. Next, in Eq.~\eqref{app:eq:trace-2}, we use the cyclic property of the trace and observe that the two terms are equal to each other and obtain the result as zero. The higher order nested commutators will consequently follow the same pattern and lead to a zero trace. Therefore, when computing the trace of a term $\mbox{Tr} (\partial_{\theta_i} e^{-\Ht} )$, it is sufficient to insert only the leading term of the series expansion, which reads,
\begin{equation}
    \mbox{Tr} (\partial_{\theta_i} e^{-\Ht} ) = -\mbox{Tr} (e^{-\Ht} H_i).
    \label{app:eq:trace-derivative}
\end{equation}

\subsection{\label{app:derivative-cost} Derivative of quantum relative entropy}

Let us start by recalling some of the definitions. Quantum relative entropy between the target ($\eta$) and model ($\rho$) density matrices is given as
\begin{equation}
    S(\eta \, || \, \rho) = \mbox{Tr}(\eta \log \eta) - \mbox{Tr}(\eta \log \rho),
\label{eq:app-quantum-rel-entropy}
\end{equation}
where both states satisfy $\mbox{Tr}(\eta) =  \mbox{Tr}(\rho) = 1$. The model density matrix is the Gibbs state of the Hamiltonian $\Ht$ at inverse temperature $\beta$, which is defined as
\begin{equation}
\rho = \frac{e^{-\beta \Ht}}{\mbox{Tr}(e^{-\beta \Ht})},
\end{equation}
where the Hamiltonian can be defined as
\begin{equation}
    \Ht = \sum_i \theta_i H_i,
\end{equation}
where $\forall\, i, \theta_i \in \mathbb{R}$ and $\mathbf{\theta}$ is the parameter vector that parametrizes each $H_i$, which are Pauli strings with length $n$ excluding the identity ($I^{\otimes n}$).

For simplicity, we set the inverse temperature $\beta$ to 1 for all derivations and experiments. Then, the derivative of the quantum relative entropy with respect to the $\theta_i$ parameter reads, 
\begin{align}
    \partial_{\theta_i} S(\eta||\rho) &= - \partial_{\theta_i} \mbox{Tr} \left(\eta \,  \mbox{log} \frac{e^{-\Ht}}{\mbox{Tr} (e^{-\Ht})}\right) \label{app:eq:derivative-1}\\
    &= - \partial_{\theta_i} \mbox{Tr} (\eta \,  (-\Ht - \mbox{log} \, \mbox{Tr} (e^{-\Ht}))) \label{app:eq:derivative-2}\\
    &= \partial_{\theta_i} \mbox{Tr} (\eta \, \Ht) + \partial_{\theta_i} \mbox{Tr} (\eta \, \mbox{log} \, \mbox{Tr} (e^{-\Ht})) \label{app:eq:derivative-3}\\
    &= \partial_{\theta_i} \mbox{Tr} (\eta \, \Ht) + \partial_{\theta_i} \mbox{Tr} (\eta) \mbox{log} \, \mbox{Tr} (e^{-\Ht}) \label{app:eq:derivative-4}\\
    &= \partial_{\theta_i} \mbox{Tr} (\eta \, \Ht) + \partial_{\theta_i} \mbox{log} \, \mbox{Tr} (e^{-\Ht}) \label{app:eq:derivative-5}\\
    &= \mbox{Tr} (\eta \, H_i) + \frac{\mbox{Tr} (\partial_{\theta_i} (e^{-\Ht}))}{\mbox{Tr} (e^{-\Ht})}  \label{app:eq:derivative-6}\\
    &= \mbox{Tr} (\eta \, H_i) - \mbox{Tr} \left( \frac{e^{-\Ht}}{\mbox{Tr} (e^{-\Ht})} H_i \right) \label{app:eq:derivative-7}\\
    &= \mbox{Tr} (\eta \, H_i) - \mbox{Tr} (\rho H_i).
\end{align}

In Eq.~\eqref{app:eq:derivative-4} we use the fact that the logarithm of the trace is a scalar and can be moved out of the trace. In Eq.~\eqref{app:eq:derivative-5}, we use the $\mbox{Tr}(\eta)=1$ property of states. In Eq.~\eqref{app:eq:derivative-7}, we insert trace of the derivative we derived in Eq.~\eqref{app:eq:trace-derivative}. Finally, we obtain the gradient as the difference of the expectation values measured on the target state and the model state.

\section{More details on numerical results}

\subsection{\label{app:tpq-sweep} Numerical errors of the TPQ states and Lanczos methods}

\begin{figure}[!h]
    \centering
\begin{minipage}{0.49\textwidth}
    \subfloat[][$n=8$ qubit case] {\includegraphics[width=\linewidth]{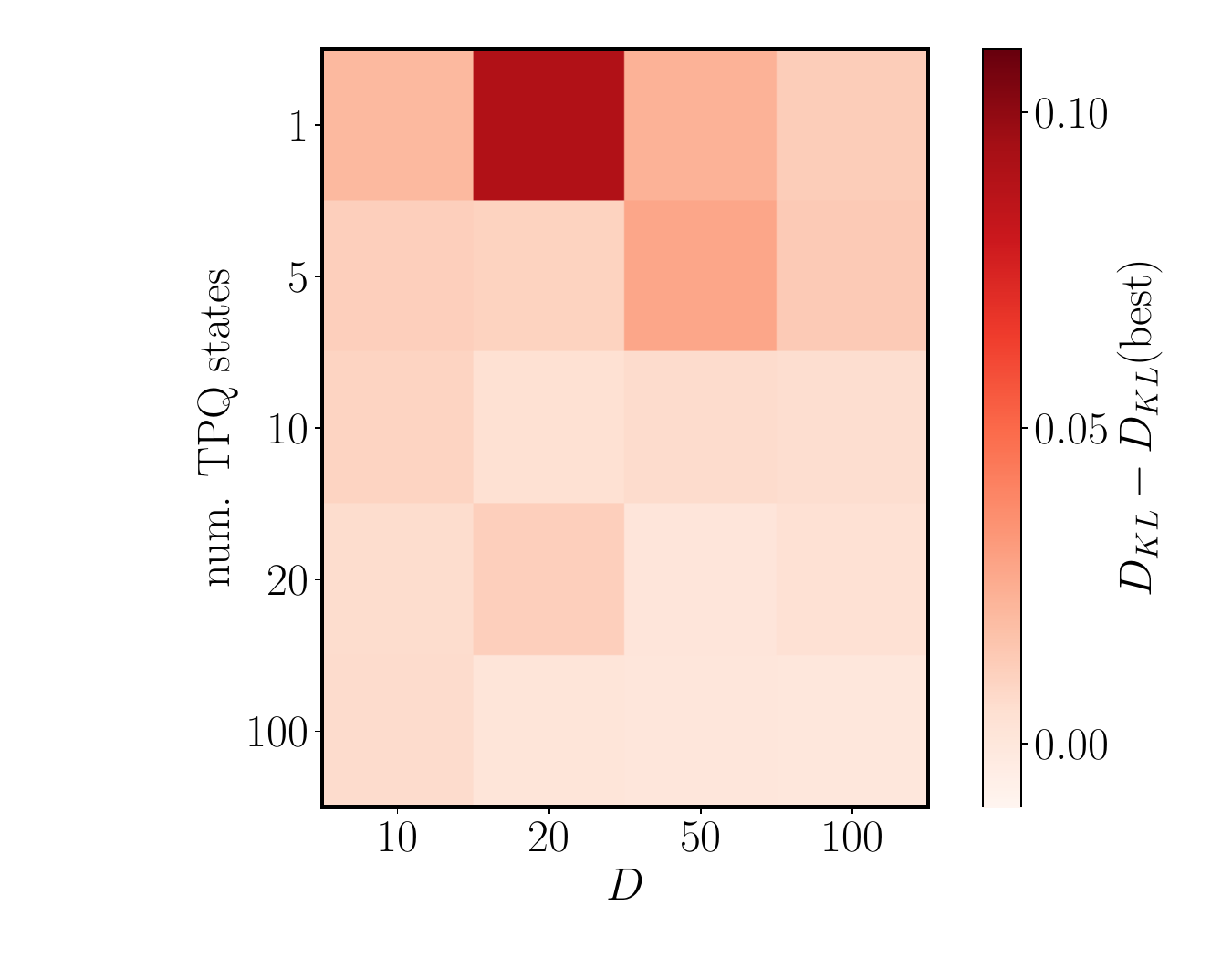}    \label{fig:tpq-8}}%
\end{minipage}
\hfill%
\begin{minipage}{0.49\textwidth}
    \subfloat[][$n=10$ qubit case] {\includegraphics[width=\linewidth]{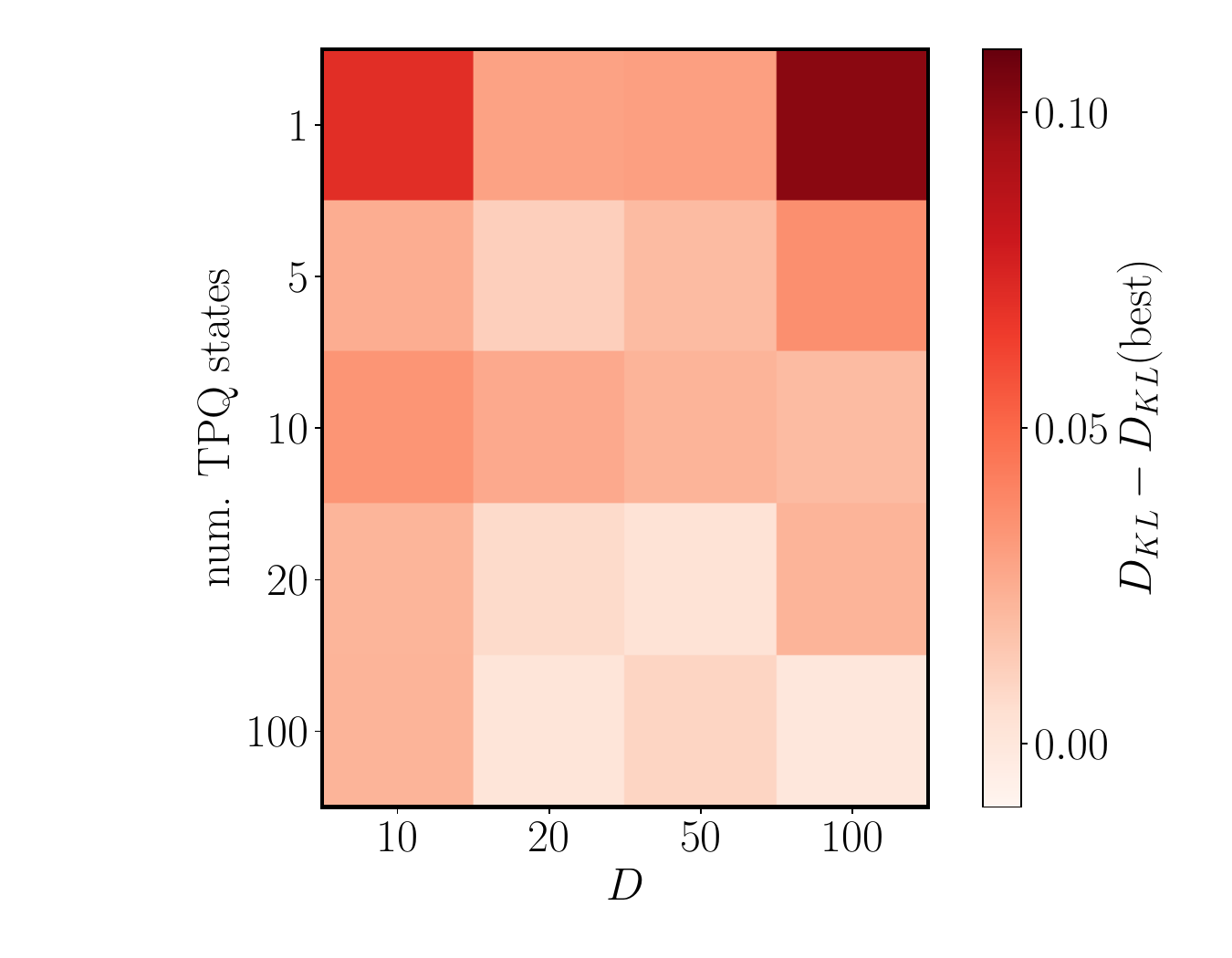}    \label{fig:tpq-10}}%
\end{minipage}
    \caption{\textbf{Model performance change with respect to different number of TPQ states and $D$ Krylov dimension for the Lanczos method.} (\textbf{a}) $\nbins = 16$, $m=2$ particles, $n=8$ qubits. (\textbf{b}) $\nbins = 32$, $m=2$ particles, $n=10$ qubits. The model is the all-to-all connected QBM with \textit{generic} Hamiltonian.}
    \label{fig:tpq-sweep}
\end{figure}

Here, we provide a study of the numerical errors on the training performance. For this purpose, we consider two target distributions from the particle physics dataset: $\nbins = 16$, $m=2$ particles, $n=8$ qubits and, $\nbins = 32$, $m=2$ particles, $n=10$ qubits. We consider the all-to-all connected QBM with \textit{generic} Hamiltonian. All models are trained using the number of TPQ states and Krylov dimension $D$ specified in Fig.~\ref{fig:tpq-sweep}. To separate the systematic errors of estimating the output distribution, the Gibbs states of all models are prepared using exact methods and $\KL$ is measured using the exact method, such that we measure only the training performance. We compare the $\KL$ values obtained using the TPQ states method to the $\KL$ value obtained using exact diagonalization. As expected, we observe that $\KL$ values get closer to the $\KL$ values of the exact diagonalization as the number of TPQ states and the Krylov dimension increases. After this study, we conclude that choosing 100 TPQ states and $D=20$ is sufficient to conduct all of the experiments. Although these values are sufficient for system sizes of $n=8$ and $n=10$ here, they will naturally result in larger errors as the system size increases.

\subsection{Connectivity definitions}

\begin{figure}[!h]
    \centering
\begin{minipage}{0.39\textwidth}
    \subfloat[All-to-all]{\includegraphics[width=.5\linewidth]{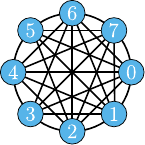}\label{fig:8q-all-to-all}}%
    
    \subfloat[][Nearest-neighbor (NN)]{
    \includegraphics[width=\linewidth]{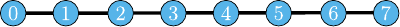}\label{fig:8q-nn}}%
    
    \subfloat[][Next-nearest-neighbor (next NN)]{\includegraphics[width=\linewidth]{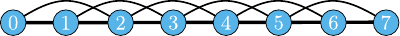}\label{fig:8q-next-nn}}%
\end{minipage}
\hfill%
\begin{minipage}{0.29\textwidth}
    \subfloat[][All-to-all]{\includegraphics[width=.9\linewidth]{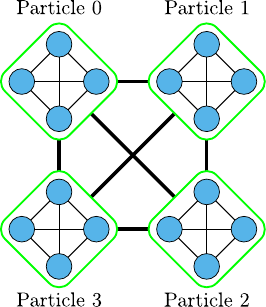}\label{fig:conn-all-to-all}}%

\end{minipage}
\hfill%
\begin{minipage}{0.29\textwidth}
    \subfloat[][Nearest-neighbor (NN)-particle]{\includegraphics[width=.9\linewidth]{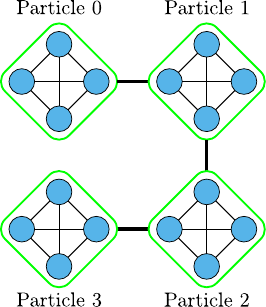}\label{fig:conn-nn-particle}}%
\end{minipage}

    \caption{\textbf{Connectivity graphs of various models.} (\textbf{a,b,c}) Connectivity graph used in eight qubit experiments. (\textbf{d, e}) Connectivity graphs of all-to-all and NN-particle (nearest-neighbor particle) cases for the $\nbins=16$ case ($n=4$ qubits for each particle). Notice that qubits that belong to a single particle are always all-to-all connected.}
    \label{fig:connectivity}
\end{figure}

In this section, we provide illustrations of different connectivity layouts used throughout this work. Fig.~\ref{fig:8q-all-to-all} shows all-to-all connectivity for $n=8$ qubits. Fig.~\ref{fig:8q-nn} shows the nearest-neighbor (NN) connectivity with open boundary conditions for $n=8$ qubits, while Fig.~\ref{fig:8q-next-nn} shows the next-nearest-neighbor connectivity with open boundary conditions for the same system size. Next, in Fig.~\ref{fig:conn-all-to-all} we illustrate all-to-all connectivity of an $n=16$ qubit system. This illustration highlights the groups of qubits that are belonging to the same particle. Since this system is meant to represent $m=4$ particles with $\nbins = 16$ each, each particle requires $n=4$ qubits ($\mbox{log}_2 16$). In Fig.~\ref{fig:conn-nn-particle}, we illustrate the NN-particle connectivity for the same system. This type of connectivity combines groups of qubits that belong to the neighboring particles with open boundary conditions. In this setting, a qubit that belongs to particle 0 is connected to $3$ (particle $0$) $+$ $4$ (particle $1$) $=$ $7$ qubits, while a qubit that belongs to particle 1 is connected to $4$ (particle $0$) $+$ $3$ (particle $1$) $+$ $4$ (particle $2$) $=$ $11$ qubits.

\subsection{Mutual information of the jet event generation problem}

\begin{figure}[!ht]
    \centering
    \includegraphics[width=.5\linewidth]{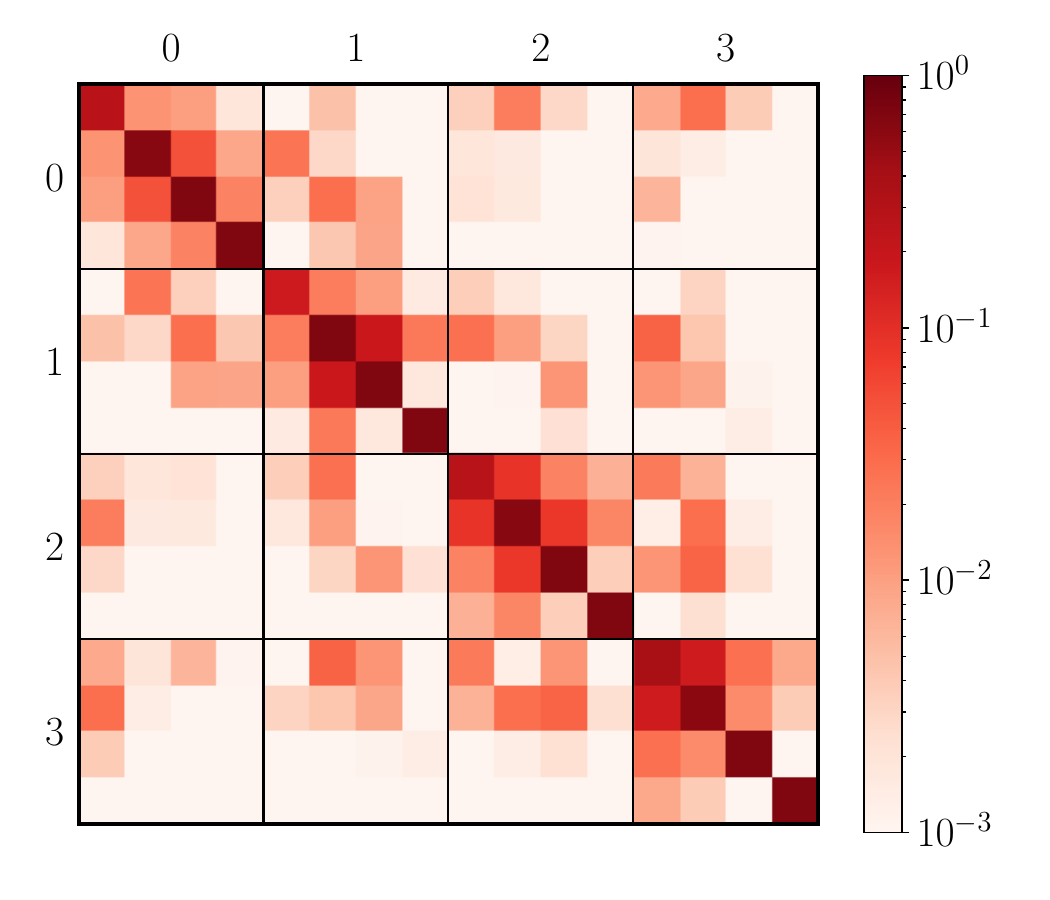}
    \caption{\textbf{Mutual information computed on the target distribution.} We consider the $m=4$ particle and $\nbins=16$ case ($n=16$ qubits).}
    \label{fig:mutinfo-target}
\end{figure}

We measure the classical mutual information on the target distribution that belongs to the $m=4$ particle and $\nbins=16$ case ($n=16$ qubits). This results in each particle being expressed with $n=4$ qubits. We measure the mutual information by resampling the \textit{train} distribution 100,000 times and using the $mutual\_info\_classif$ function of \textsc{Scikit-learn}~\cite{scikit-learn}. We present all pair-wise mutual information values in Fig.~\ref{fig:mutinfo-target}. We observe non-zero values that connect at least one unit from each particle with each other and almost all units are connected to others in no particular order. As expected, units that correspond to the same particle have more mutual information.

\subsection{\label{app:ham-comp} Effect of model Hamiltonian and connectivity to QBM expressivity}

\begin{figure*}[!ht]
    \centering
    \subfloat[][$m=2$ particles]{\includegraphics[width=.33\textwidth]{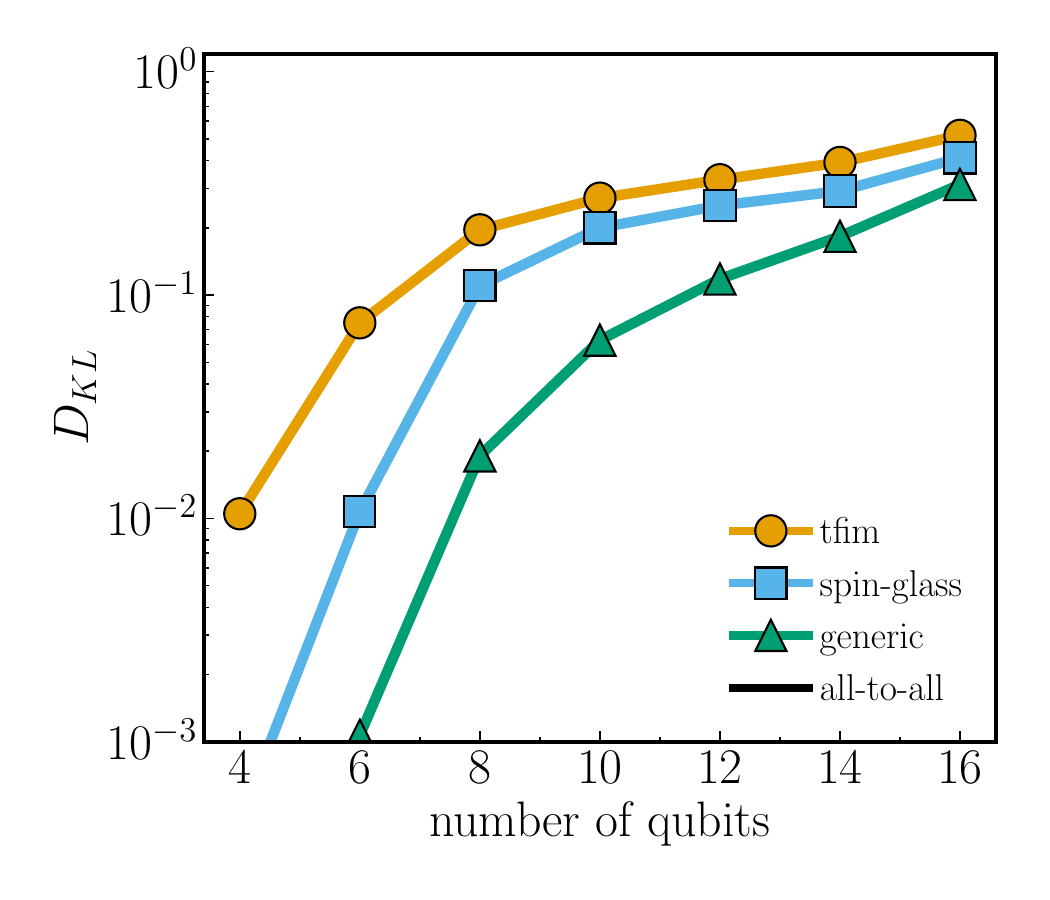}\label{fig:conn-2}}%
    \subfloat[][$m=3$ particles]{\includegraphics[width=.33\textwidth]{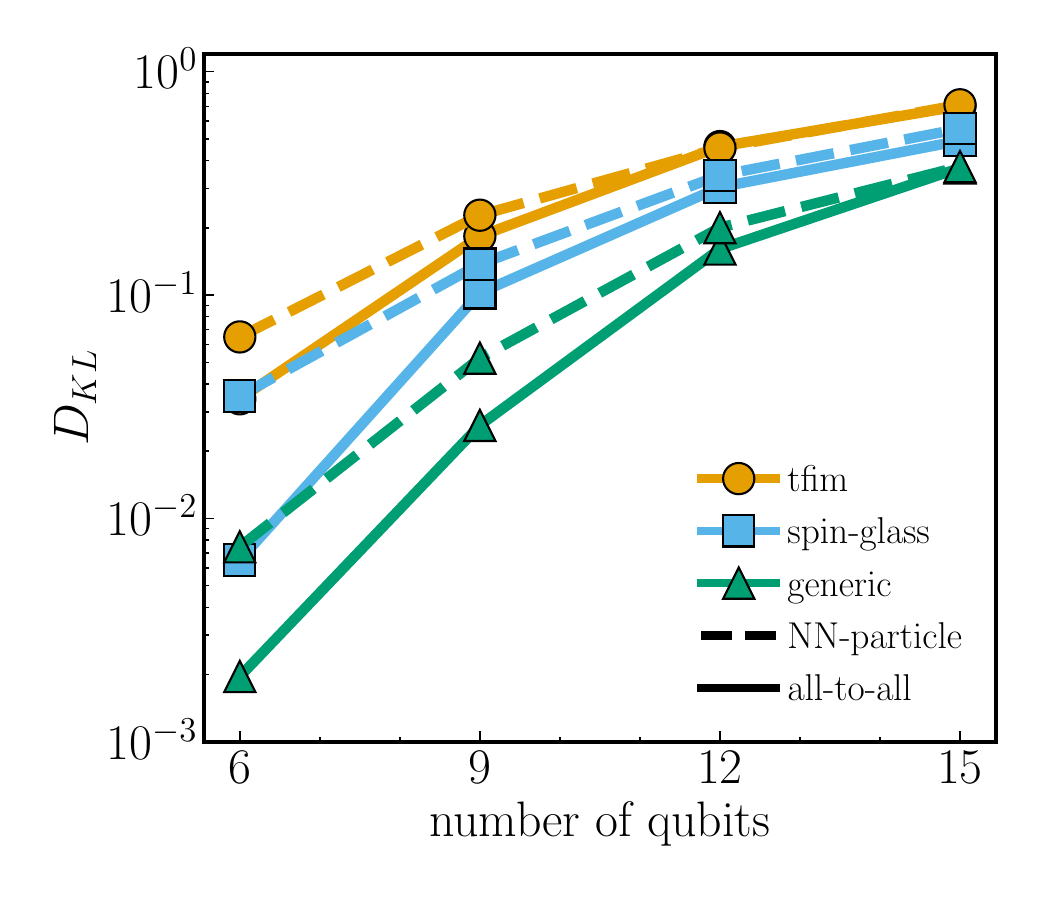}\label{fig:conn-3}}%
    \subfloat[][$m=4$ particles]{\includegraphics[width=.33\textwidth]{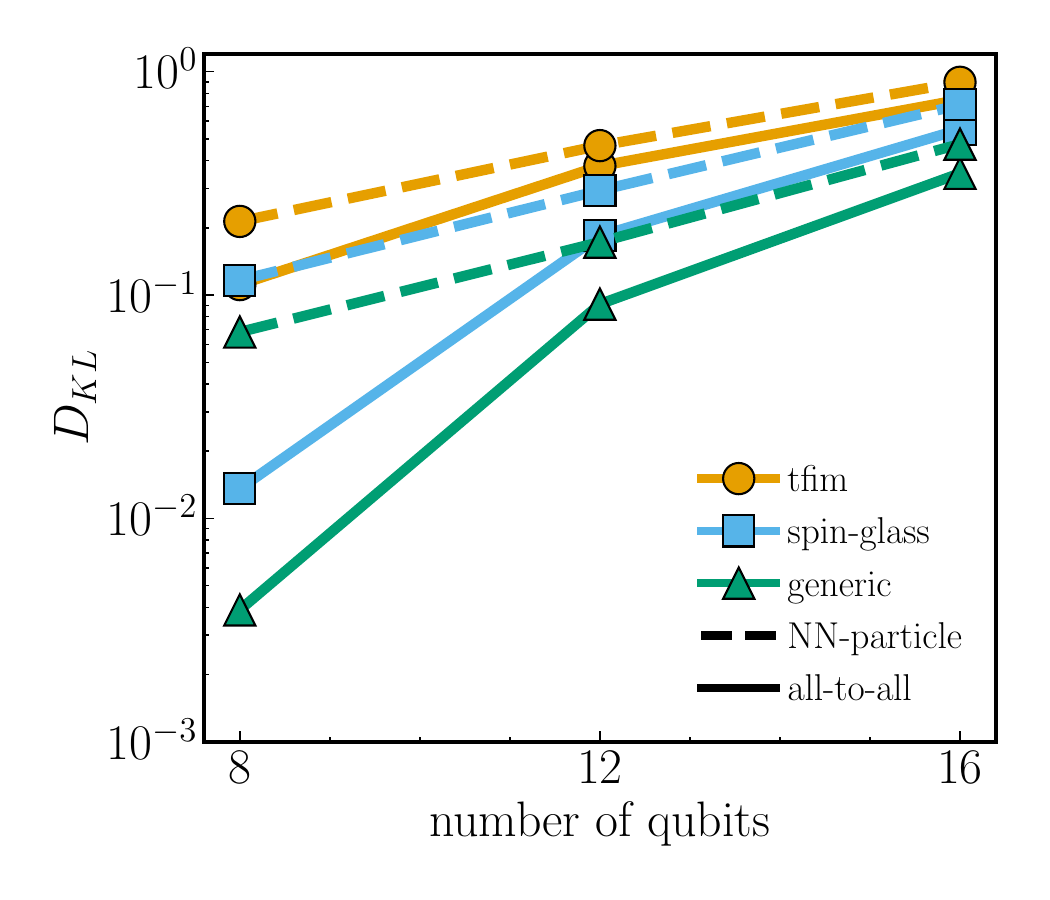}\label{fig:conn-4}}%
    \caption{\textbf{Comparison of QBMs with different Hamiltonians and connectivity.} Solid lines refer to all-to-all connectivity, while dashed lines refer to NN-particle connectivity. All-to-all and NN-particle cases are equivalent for the $m=2$ particle case.}
    \label{fig:hamiltonian-type}
\end{figure*}

We compare the effect of the model Hamiltonian to QBM expressivity by considering three types of Hamiltonians and two types of connectivity. The Hamiltonian definitions are provided in Table~\ref{tab:hamiltonian-defs} and connectivity definitions are provided in Fig.~\ref{fig:conn-all-to-all} and Fig.~\ref{fig:conn-nn-particle}. All models are trained using TPQ states and the output probability is approximated using the Lanczos method. $\KL$ values measured after training for three different cases are provided in Fig.~\ref{fig:hamiltonian-type}. We observe that the QBM is able to learn the target distributions better as the Hamiltonian contains more terms. We also observe that a Hamiltonian with more terms but with fewer connections can get better results than a Hamiltonian with fewer terms but more connections. This points to the fact that both Hamiltonian terms and connectivity are resources that contribute to the expressive power of the model in different ways.

\subsection{\label{app:sec:temperature} Effect of inverse temperature to model performance after training}

Here, we test the impact of inverse temperature after training. Each model is technically trained to an effective temperature. During training we set $\beta=1$, however the coefficients are not constrained. Therefore, their absolute values can go beyond 1.0. This leads to a change in the effective inverse temperature. Recall the definition from the main text for the effective temperature $\Tilde{\beta}$:
\begin{equation}
    \Tilde{\beta} = \mbox{max} \left( | \mathbf{\theta}| \right).
\end{equation}

\begin{figure}[!ht]
    \centering
    \begin{minipage}{0.49\textwidth}
        \subfloat[][$n=8$ qubit case] {\includegraphics[width=\linewidth]{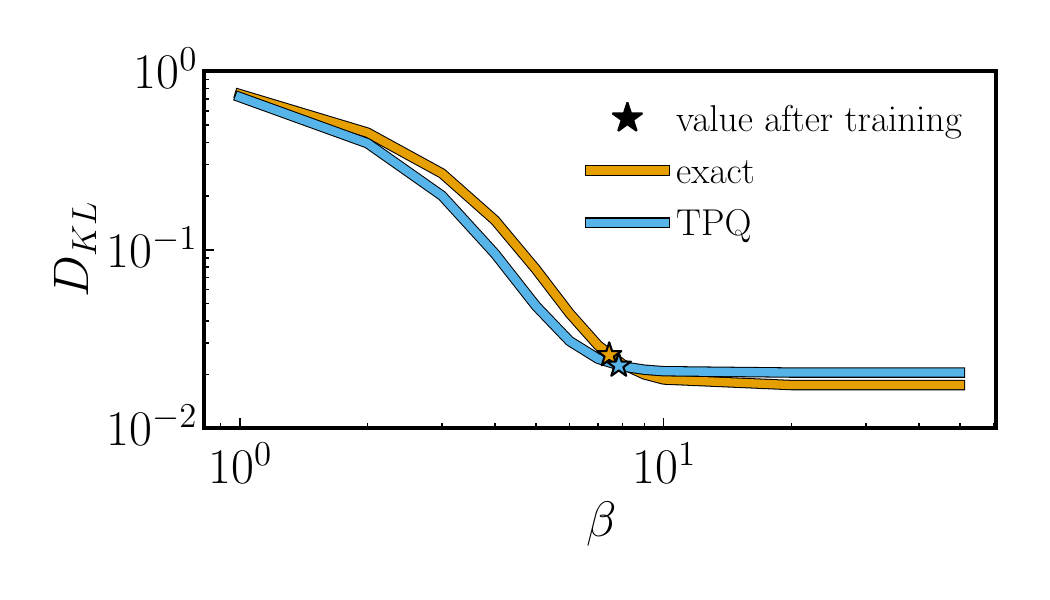}    \label{fig:beta-test-1}}%
    \end{minipage}
    \hfill%
    \begin{minipage}{0.49\textwidth}
        \subfloat[][$n=10$ qubit case] {\includegraphics[width=\linewidth]{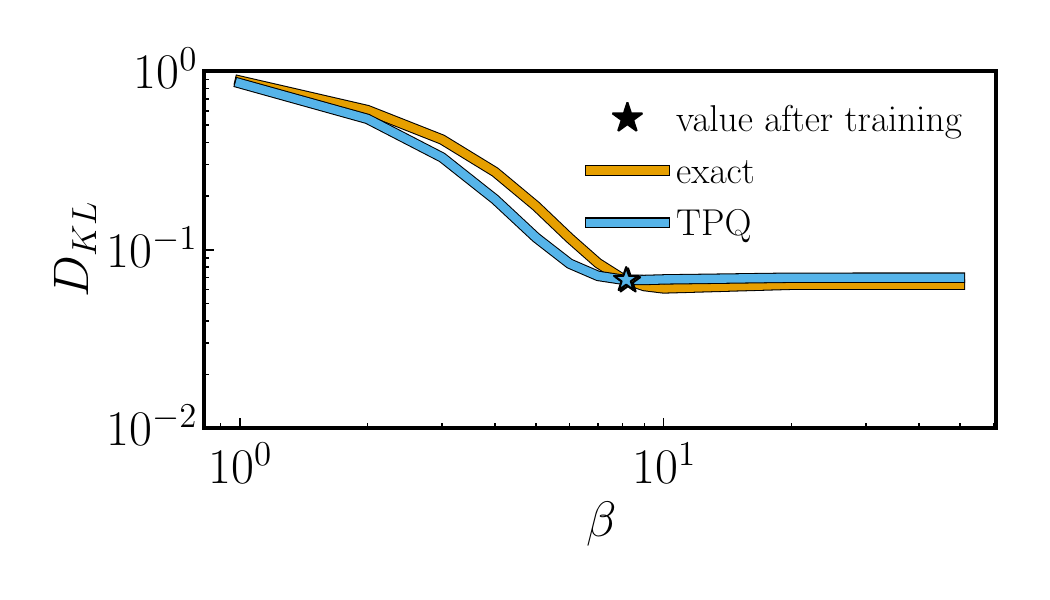}   \label{fig:beta-test-2}}%
    \end{minipage}
    \caption{\textbf{Model performance change with respect to $\beta$ inverse temperature.} (\textbf{a}) $\nbins = 16$, $m=2$ particles, $n=8$ qubits. (\textbf{b}) $\nbins = 32$, $m=2$ particles, $n=8$ qubits. The model is the all-to-all connected QBM with \textit{generic} Hamiltonian. Stars denote $\Tilde{\beta}$ reached at the end of training. We evaluate the model on different values of the effective temperature.}
    \label{fig:beta-test}
\end{figure}

We vary $\Tilde{\beta}$ and observe how it impacts the output probability distribution. We present results for the cases of $\nbins=16$ (Fig.~\ref{fig:beta-test-1}) and $\nbins=32$ (Fig.~\ref{fig:beta-test-2}) with $m=2$ particles as the target distributions. We train both models using exact diagonalization as well as the TPQ states method. We evaluate $\KL$ using the exact diagonalization in order to isolate the systematic errors. We observe that the models find a close-to-optimal value for $\Tilde{\beta}$ after training. Increasing its value does not significantly change $\KL$, but decreasing its value only results in worse $\KL$ values.
 
\end{document}